# Error Correcting Coding for a Non-symmetric Ternary Channel

Nicolas Bitouzé, *Student Member, IEEE,* Alexandre Graell i Amat, *Member, IEEE,*
and Eirik Rosnes, *Member, IEEE*

*Abstract*—Ternary channels can be used to model the behavior of some memory devices, where information is stored in three different levels. In this paper, error correcting coding for a ternary channel where some of the error transitions are not allowed, is considered. The resulting channel is non-symmetric, therefore classical linear codes are not optimal for this channel. We define the maximum-likelihood (ML) decoding rule for ternary codes over this channel and show that it is complex to compute, since it depends on the channel error probability. A simpler alternative decoding rule which depends only on code properties, called $d_A$-decoding, is then proposed. It is shown that $d_A$-decoding and ML decoding are equivalent, i.e., $d_A$-decoding is optimal, under certain conditions. Assuming $d_A$-decoding, we characterize the error correcting capabilities of ternary codes over the non-symmetric ternary channel. We also derive an upper bound and a constructive lower bound on the size of codes, given the code length and the minimum distance. The results arising from the constructive lower bound are then compared, for short sizes, to optimal codes (in terms of code size) found by a clique-based search. It is shown that the proposed construction method gives good codes, and that in some cases the codes are optimal.

## I. Introduction

Error correcting coding plays a central role in any communication system. Since the seminal paper by Shannon, the main body of research on coding theory has been devoted to binary linear codes. However, non-binary codes have also demonstrated remarkable performance. Among them, Reed-Solomon codes [1] are one of the most popular and widely used coding schemes. Recently, the interest for non-binary codes has been renewed with the rediscovery of low-density parity-check (LDPC) codes [2]. Non-binary LDPC codes have been shown to perform very close to capacity and to outperform binary LDPC codes in some cases [3].

Most of the previous works on non-binary codes consider a Galois Field whose order $q$ is a power of 2. On the other hand, little attention has been devoted to non-binary codes when $q$ is not a power of two. Indeed, for conventional channels, binary linear codes or $q$-ary codes with $q$ being a power of two show very good performance. Ternary codes for

The material in this paper was presented in part at the 2009 Information Theory and Applications Workshop, La Jolla, CA, February 2009.

N. Bitouzé and A. Graell i Amat are with the Department of Electronics, Institut TELECOM-TELECOM Bretagne, CS 83818 - 29238 Brest Cedex 3, France (e-mail: nicolas.bitouze@telecom-bretagne.eu,alexandre.graell@telecom-bretagne.eu). E. Rosnes is with the Selmer Center, Department of Informatics, University of Bergen, N-5020 Bergen, Norway (e-mail: eirik@ii.uib.no). This work was partially funded by a Marie Curie Intra-European Fellowship within the 6th European Community Framework Programme and by the Norwegian Research Council (NFR) under Grants 174982 and 183316.

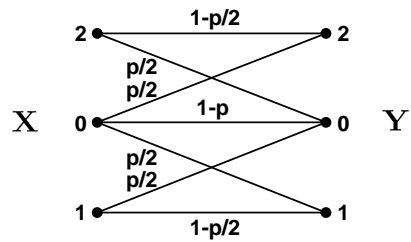

Fig. 1. Non-symmetric ternary channel.

the symmetric ternary channel using the ordinary Hamming distance metric have been considered in the literature before. See, for instance, [4, 5] and references therein. In this paper, however, we consider error correcting coding for a ternary non-conventional channel.

Recently, coding for flash memories has received some attention. See, for instance, [6–8] and references therein. Multilevel flash memory is a storage technology where the charge level of any cell can be easily increased, but not easily decreased. In fact, the only way to decrease the charge level of a cell is to erase the whole block (i.e., set the charge on all cells in a block to zero) and reprogram each cell. This is a time-consuming process which consumes energy and reduces the lifetime of the memory. The coding problem for flash memories is to design modulation codes that maximize the number of rewrites between two erasures.

In this paper, however, we look at a different memory device coding problem, namely coding for electrically erasable programmable read-only memories (EEPROMs), which are semiconductor memories that retain their data contents when power is off. They can be read and written to like standard RAMs and are suitable for applications where storage of small amounts of data is critical and periodic writing of new data is required. Typical applications are radio frequency identification tag, smart dust, or automotive applications including car audio and multimedia, chassis and safety, and power train. The communication channel underlying EEPROMs can be suitably modeled as a binary symmetric channel (BSC). Currently, very simple error correcting codes based on the well-known Hamming codes combined with hard decoding are implemented on-chip to correct single bit errors [9]. However, next generation devices demand for more stringent requirements in terms of reliability as well as storage density. A suitable modification of the physics of EEPROM memories allows to store the information in three levels, thus higher densities can be achieved. While transitions between adjacent levels are allowed, transitions from the highest level to the lowest level and from the lowest level to the highest level are physically not possible. A simple model for the resulting channel is the discrete memoryless ternary channel with input alphabet $\mathcal{X} = \{0,1,2\}$, output alphabet $\mathcal{Y} = \{0,1,2\}$, and probability transition matrix

$$p(y|x) = \begin{pmatrix} 1-p & p/2 & p/2 \\ p/2 & 1-p/2 & 0 \\ p/2 & 0 & 1-p/2 \end{pmatrix} \quad (1)$$

where $p \leq 2/3$ and the entry in the $i$th row and the $j$th column denotes the conditional probability of receiving symbol $j$ when symbol $i$ was transmitted. Notice that transitions $1 \to 2$ and $2 \to 1$ are not allowed. As a result, the channel defined by (1) is non-symmetric. The channel model is depicted in Fig. 1.

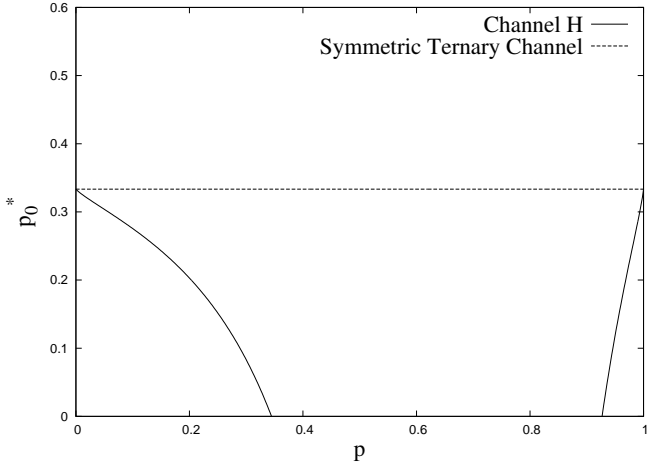

Fig. 2. Ratio of symbols 0 in the optimal input distribution $p(x)$ that maximizes $I(X,Y)$ of channel $\mathcal{H}$ as a function of $p$.

In this paper, we consider error correcting coding for the non-symmetric ternary channel of Fig. 1. We define the maximum-likelihood (ML) decoding rule for ternary codes over this channel and show that its implementation is complex, since it depends on the channel error probability $p$. As an alternative, a simpler decoding rule which depends only on code properties, called $d_A$-decoding, is proposed based on a more appropriate distance measure. It is shown that under certain conditions the proposed decoding rule is optimal, i.e., it is equivalent to ML decoding. We then address error correcting capabilities of ternary codes under $d_A$-decoding. In particular, we derive a sphere-packing bound to upper bound the size of the codes assuming $d_A$-decoding. We also derive a constructive lower bound on the size of the codes given the code length and its minimum distance, which proves the existence of good codes. The construction method is based on binary block codes as basic elements. The construction method is then generalized to a non-symmetric $q$-ary channel. Finally, for ternary codes of small sizes, we compare the constructive lower bounds to optimal codes (in terms of code size) found by a clique-based code search. It is shown that the binary code construction method gives very good codes. Also, in some cases, optimal ternary codes are obtained.

The remainder of the paper is organized as follows. In Section II, we outline some of the notation used in this paper. Then, in Section III, we address the computation of the channel capacity. The ML decoding rule for ternary codes over the non-symmetric channel is given in Section IV. Also, a simpler decoding rule which depends only on code properties is derived. Section V addresses the error correcting capabilities of ternary codes over the non-symmetric ternary channel. In Section VI, an upper bound on the size of codes is given, and in Section VII a constructive lower bound is derived. An encoding algorithm is given in Section VIII. The construction method is generalized to $q$-ary non-symmetric channels in Section IX, and in Section X we compare the values from the constructive lower bound with the results of a clique-based search. Finally, in Section XI, we draw some conclusions.

## II. NOTATION

Throughout the paper, we use capital letters to denote random variables, e.g., $X$, and its calligraphic version, $\mathcal{X}$, to denote the alphabet of $X$. Also, for convenience, we denote the probability mass function by $p(x) = \Pr(X = x), x \in \mathcal{X}$, rather than by $p_X(x)$. We will write vectors in boldface letters, and the $i$th element of a vector $\boldsymbol{a}$ as $a_i$. The cardinality of a set $\mathcal{S}$, i.e., the number of elements in $\mathcal{S}$ is denoted by $|\mathcal{S}|$. Furthermore, a $q$-ary code $\mathcal{C}$ of length $n$ is a subset of $\mathbb{Z}_q^n$, $q \geq 2$, where $\mathbb{Z}_q = \{0, \ldots, q-1\}$ and $\mathbb{Z}_q^n$ is the set of all $n$-tuples over $\mathbb{Z}_q$. We will use subindexes to distinguish between codes over alphabets of different order $q$. For instance, a binary code will be denoted by $\mathcal{C}_2$ and a ternary code by $\mathcal{C}_3$. A code $\mathcal{C}$ with code length $n$ containing $M = |\mathcal{C}|$ codewords and minimum distance $d$ shall be referred to as an $[n, M, d]$ code. The Hamming weight of a vector $\boldsymbol{a}$ is denoted by $w_{\boldsymbol{a}}$ and the Hamming distance between two vectors $\boldsymbol{a}$ and $\boldsymbol{b}$ is denoted by $d_H(\boldsymbol{a}, \boldsymbol{b})$. For simplicity, we shall denote the non-symmetric ternary channel of Fig. 1 by $\mathcal{H}$.

## III. CHANNEL CAPACITY

In this Section, we derive the capacity $(C)$ for the channel model of Fig. 1, defined as

$$C \triangleq \max_{p(x)} I(X, Y) \quad (2)$$

where $I(X, Y)$ is the mutual information between $X$ and $Y$. We denote by $p_x$, $x \in \{0, 1, 2\}$, the probability $\Pr(X = x)$. Due to the symmetry between symbols 1 and 2, we can assume that the input distribution $p(x) = \langle p_0, p_1, p_2 \rangle$ which maximizes $I(X, Y)$ will be such that $p_1 = p_2$. Thus, the distributions we are interested in are entirely characterized by $p_0$ and take the form $p(x) = \langle p_0, \frac{1-p_0}{2}, \frac{1-p_0}{2} \rangle$. With this constraint, $I(X, Y)$ can be written as

$$I(X,Y) = h\left(p_0 + \frac{p}{2} - \frac{3}{2}p_0 p\right) - p_0 h(p) - (1-p_0)h\left(\frac{p}{2}\right) \\ + (1-p_0)\left(1 - \frac{p}{2}\right)\log_3(2) \quad (3)$$

where $h(t) = -t\log_3(t) - (1-t)\log_3(1-t)$ is the ternary entropy function.

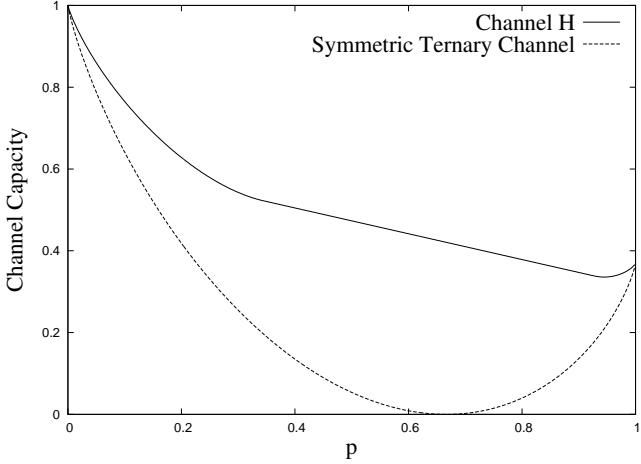

Fig. 3. Channel capacity of the ternary channel of Fig. 1 and of the symmetric ternary channel as a function of the error probability $p$.

Let $f$ be the function $f = \frac{\partial I(X,Y)}{\partial p_0}$, the partial derivative of $I(X,Y)$ with respect to $p_0$. The zeros of the function $f$ are the values of $p_0$ that maximize $I(X,Y)$. We denote the values of $p_0$ which maximize $I(X,Y)$ by $p_0^*$. $p_0^*$ can be written as

$$p_0^* = \frac{\frac{3^{\lambda(p)}}{1+3^{\lambda(p)}} - \frac{p}{2}}{1 - \frac{3}{2}p} \quad (4)$$

where

$$\lambda(p) = \frac{h(p/2) - h(p) - \left(1 - \frac{p}{2}\right)\log_3(2)}{1 - \frac{3}{2}p}. \quad (5)$$

The values of $p_0^*$ are given in Fig. 3 as a function of $p$. Since $\mathcal{H}$ is not symmetric, the input distribution $p(x)$ that maximizes the mutual information $I(X,Y)$ is not uniform. For very low values of $p$, the best input distribution tends to the uniform distribution. However, for increasing values of $p$ the optimal distribution tends to favor the symbols 1 and 2, and symbol 0 should be less used. There is a point after which the best distribution is $p(x) = \langle 0, 1/2, 1/2 \rangle$ for some range of values of $p$. This implies that symbol 0 should not be used for this range of transition error probabilities. In that case, the optimal codes are binary codes on symbols $\{1, 2\}$. For values of $p$ approaching one, the best distribution tends again to the uniform distribution.

The channel capacity is depicted in Fig. 3 as a function of $p$. For comparison purposes, the capacity of the symmetric ternary channel is also given. Clearly, the capacity of the non-symmetric ternary channel is higher.

## IV. ML Decoding and $d_A$-decoding

In this Section, we give the ML decoding rule for the ternary channel $\mathcal{H}$ of Fig. 1. We then propose an alternative decoding rule, called $d_A$-decoding, which is much simpler to compute, and show that both rules are equivalent under certain conditions.

For later use, let $\mathcal{C}_3 \subset \mathbb{Z}_3^n$ be a ternary code of length $n$. Also, let $\mathbf{x} = (x_1, \ldots, x_n)$ be a codeword in $\mathcal{C}_3$ which is transmitted over channel $\mathcal{H}$, and $\mathbf{y} = (y_1, \ldots, y_n) \in \mathbb{Z}_3^n$ the received noisy observation at the output of the channel. The user data is assumed to be uniformly distributed, and thus also the codewords.

Let $\mathbf{u}, \mathbf{v} \in \mathbb{Z}_3^n$ be two ternary vectors, and define the subsets $\mathcal{S}_0$, $\mathcal{S}_1$, $\mathcal{S}_2$, and $\mathcal{S}_3$ as

$$\begin{aligned} \mathcal{S}_0 &= \{i : u_i = v_i = 0\} \\ \mathcal{S}_1 &= \{i : u_i = v_i \neq 0\} \\ \mathcal{S}_2 &= \{i : u_i \neq v_i \wedge u_i v_i = 0\} \\ \mathcal{S}_3 &= \{i : u_i \neq v_i \wedge u_i v_i \neq 0\}. \end{aligned} \quad (6)$$

We define the following distance measure between two ternary vectors $\mathbf{u}$ and $\mathbf{v}$ transmitted over channel $\mathcal{H}$:

*Definition 1:* Let $\mathbf{u} = (u_1, \ldots, u_n)$ and $\mathbf{v} = (v_1, \ldots, v_n)$ be two vectors in $\mathbb{Z}_3^n$ transmitted over channel $\mathcal{H}$ with error probability $p$. The distance $d_{\mathrm{ML}}(\mathbf{u}, \mathbf{v})$ between $\mathbf{u}$ and $\mathbf{v}$ is defined as

$$d_{\mathrm{ML}}(\mathbf{u}, \mathbf{v}) = \begin{cases} \infty & \text{if } |\mathcal{S}_3| > 0 \\ \begin{array}{l} -|\mathcal{S}_0|\log(1-p) \\ -|\mathcal{S}_1|\log(1-p/2) \\ -|\mathcal{S}_2|\log(p/2) \end{array} & \text{otherwise} \end{cases} \quad (7)$$

*Remark 1:* Notice that with some abuse of language, we call $d_{\mathrm{ML}}$ a distance measure. However, formally speaking $d_{\mathrm{ML}}$ is not a distance measure, since the identity of indiscernibles does not hold. Also, note that for symmetric channels the distinction between subsets $\mathcal{S}_0$ and $\mathcal{S}_1$ is not necessary, since the conditional probabilities $p(y|x)$ are independent of $x$. Similarly, the distinction between subsets $\mathcal{S}_2$ and $\mathcal{S}_3$ is not required for symmetric channels.

We can express the ML decoding rule

$$\hat{\mathbf{x}} = \underset{\mathbf{x} \in \mathcal{C}_3}{\operatorname{argmax}} \, p(\mathbf{y}|\mathbf{x}) \quad (8)$$

in terms of the distance $d_{\mathrm{ML}}(\mathbf{x}, \mathbf{y})$. By taking the logarithm of the conditional probability $p(\mathbf{y}|\mathbf{x})$ we obtain:

$$-\log(p(\mathbf{y}|\mathbf{x})) = \sum_{i=1}^{n} -\log(p(y_i|x_i)) = d_{\mathrm{ML}}(\mathbf{x}, \mathbf{y}) \quad (9)$$

where the first equality is due to the assumption that the channel is memoryless. Using (9), the ML decoding rule can then be formulated as follows:

*Given a received word $\mathbf{y}$, decode to the codeword $\mathbf{x}$ that minimizes the distance $d_{\mathrm{ML}}(\mathbf{x}, \mathbf{y})$.*

*Proof:* It is sufficient to prove that for a given $\mathbf{y}$, when $\mathbf{x}$ varies among codewords, $d_{\mathrm{ML}}(\mathbf{x}, \mathbf{y})$ increases for decreasing values of $p(\mathbf{y}|\mathbf{x})$. For $d_{\mathrm{ML}}(\mathbf{x}, \mathbf{y}) = \infty$ there is at least one position $i$ such that the transition $x_i \to y_i$ is not permitted; therefore $p(\mathbf{y}|\mathbf{x}) = 0$. Now, we consider the case where $d_{\mathrm{ML}}(\mathbf{x}, \mathbf{y}) < \infty$, in which case

$$p(\mathbf{y}|\mathbf{x}) = \exp(-d_{\mathrm{ML}}(\mathbf{x}, \mathbf{y})). \quad (10)$$

Then, the result follows from the monotonicity of the exponential function. ∎

Notice that $d_{\mathrm{ML}}$ depends on both the code and the channel transition probability $p$. However, one would be interested in a distance metric that depends only on the code, thus allowing

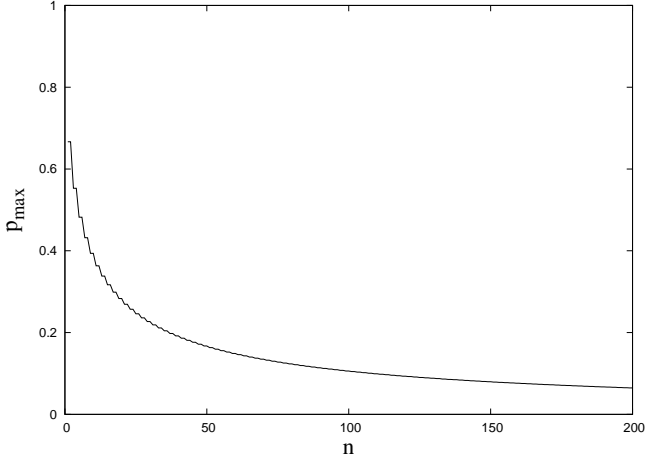

Fig. 4. Maximum value of $p$, $p_{\max}$, for the equivalence between $d_{\mathrm{A}}$-decoding and ML decoding rules as a function of code length $n$.

for a simpler decoding rule. We define the following distance measure between two ternary vectors $\mathbf{u}$ and $\mathbf{v}$:

*Definition 2:* Let $\mathbf{u}$ and $\mathbf{v}$ be two vectors in $\mathbb{Z}_3^n$. The distance $d_{\mathrm{A}}(\mathbf{u}, \mathbf{v})$ between $\mathbf{u}$ and $\mathbf{v}$ is defined as

$$d_{\mathrm{A}}(\mathbf{u}, \mathbf{v}) = \sum_{i=1}^{n} d_{\mathrm{A}}(u_i, v_i) \qquad (11)$$

where

$$d_{\mathrm{A}}(u_i, v_i) = \begin{cases} 0 & \text{if } u_i = v_i, \\ 1 & \text{if } u_i \neq v_i \wedge u_i v_i = 0, \\ \infty & \text{if } u_i \neq v_i \wedge u_i v_i \neq 0. \end{cases} \qquad (12)$$

*Remark 2:* Notice that in this case the identity of indiscernibles holds. However, the triangular inequality does not hold anymore.

Using the distance $d_{\mathrm{A}}(\mathbf{u}, \mathbf{v})$ we can define the following decoding rule which does not depend on $p$:

*Given a received word $\mathbf{y}$, decode to the codeword $\mathbf{x}$ that minimizes the distance $d_{\mathrm{A}}(\mathbf{x}, \mathbf{y})$.*

In the remainder of the paper we shall refer to this decoding rule as $d_{\mathrm{A}}$-decoding. We denote by $t_{\mathrm{A}}$ the error correcting capability of a code $\mathcal{C}_3$ over the channel $\mathcal{H}$ under $d_{\mathrm{A}}$-decoding. Note that $d_{\mathrm{A}}$-decoding does not necessarily minimize the probability of error. However, we can prove the following Theorem:

*Theorem 1:* Let $\mathcal{C}_3$ be a ternary code of length $n$, and let $\mathcal{H}$ be the ternary channel of Fig. 1 with transition error probability $p$. $d_{\mathrm{A}}$-decoding and ML decoding of codewords transmitted over $\mathcal{H}$ are equivalent for all codes $\mathcal{C}_3$ of length $n$ if and only if the following inequality is satisfied:

$$\frac{p/2}{1-p} < \left(\frac{1-p}{1-p/2}\right)^{\lfloor \frac{n-1}{2} \rfloor}. \qquad (13)$$

*Proof:* See Appendix A. ∎

Theorem 1 gives the range of values of the channel error probability $p$ such that $d_{\mathrm{A}}$-decoding is equivalent to ML decoding. We denote the maximum value of $p$ such that $d_{\mathrm{A}}$-decoding and ML decoding are equivalent by $p_{\max}$. The value $p_{\max}$ is depicted in Fig. 4 as a function of the code length $n$. $d_{\mathrm{A}}$-decoding and ML decoding are equivalent for all values of $p$ under the curve. For small values of $n$, $d_{\mathrm{A}}$-decoding and ML decoding are equivalent even for high values of $p$. The range of $p$ for which $d_{\mathrm{A}}$-decoding is optimal decreases with the block length $n$. For instance, for $n \approx 100$, $d_{\mathrm{A}}$-decoding is optimal for $p < 0.1$. This is by far compatible with requirements of memory devices. Therefore, for practical purposes, the $d_{\mathrm{A}}$-decoding rule is optimal and can be considered instead of the more complex ML decoding rule with no loss in performance.

## V. Error Correcting Capabilities

In the following, we analyze the distance properties and the error correcting capabilities of ternary codes over the ternary channel $\mathcal{H}$ under the $d_{\mathrm{A}}$-decoding rule defined in the previous Section. We require the definition of another distance measure:

*Definition 3:* Let $\mathbf{u}$ and $\mathbf{v}$ be two vectors in $\mathbb{Z}_3^n$. The distance $d_{\mathrm{B}}(\mathbf{u}, \mathbf{v})$ between $\mathbf{u}$ and $\mathbf{v}$ is defined as

$$d_{\mathrm{B}}(\mathbf{u}, \mathbf{v}) = \min_{\mathbf{w} \in \mathbb{Z}_3^n} \left( d_{\mathrm{A}}(\mathbf{u}, \mathbf{w}) + d_{\mathrm{A}}(\mathbf{w}, \mathbf{v}) \right). \qquad (14)$$

It is easy to check that $d_{\mathrm{B}}$ is such that for two ternary symbols $u_i, v_i \in \mathbb{Z}_3$ the following equalities are satisfied:

$$d_{\mathrm{B}}(u_i, v_i) = \begin{cases} 0 & \text{if } u_i = v_i, \\ 1 & \text{if } u_i \neq v_i \wedge u_i v_i = 0, \\ 2 & \text{if } u_i \neq v_i \wedge u_i v_i \neq 0. \end{cases} \qquad (15)$$

*Remark 3:* Notice that for two binary vectors $\mathbf{u}$ and $\mathbf{v}$, $d_{\mathrm{B}}(\mathbf{u}, \mathbf{v}) = d_{\mathrm{H}}(\mathbf{u}, \mathbf{v})$.

We define the minimum $d_{\mathrm{B}}$-distance of a ternary code, denoted by $d_{\mathrm{B,min}}$, as follows:

*Definition 4:* Let $\mathbf{x}$ and $\tilde{\mathbf{x}}$ be two distinct codewords of $\mathcal{C}_3$. The minimum $d_{\mathrm{B}}$-distance of code $\mathcal{C}_3$ is

$$d_{\mathrm{B,min}} = \min_{\substack{\mathbf{x}, \tilde{\mathbf{x}} \in \mathcal{C}_3 \\ \mathbf{x} \neq \tilde{\mathbf{x}}}} d_{\mathrm{B}}(\mathbf{x}, \tilde{\mathbf{x}}). \qquad (16)$$

Then, assuming $d_{\mathrm{A}}$-decoding, the error correcting capability $t_{\mathrm{A}}$ of a ternary code over the channel $\mathcal{H}$ is given by the following Proposition:

*Proposition 1:* The error correcting capability $t_{\mathrm{A}}$ of a code $\mathcal{C}_3$ over the ternary channel $\mathcal{H}$ is

$$t_{\mathrm{A}} = \left\lfloor \frac{d_{\mathrm{B,min}} - 1}{2} \right\rfloor. \qquad (17)$$

*Proof:* See Appendix B. ∎

## VI. A Sphere-packing Bound

The main goal when designing codes is that of achieving the largest possible minimum distance with the highest possible code rate. In this Section, we give a simple upper bound on the size of codes over the ternary channel $\mathcal{H}$ assuming $d_{\mathrm{A}}$-decoding. In particular, we derive a sphere-packing bound. However, its formulation is harder than for the case of symmetric channels. Since transitions $1 \to 2$ and $2 \to 1$ are not possible, the ternary space we deal with is not isotropic and has the shape of a hypercube of dimension $n$ centered on the all-zero vector (see Fig. 5 for $n = 3$). Therefore, spheres

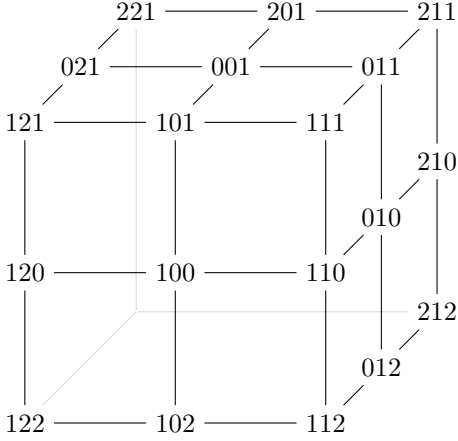

Fig. 5. $\mathbb{Z}_3^3$ with distance $d_B$.

have smaller volumes if they are closer to the vertices of the hypercube. The goal here is to find how many spheres of a given radius can be packed in the ternary space. Denote by

$$\mathcal{S}(\mathbf{u}, r) = \{\mathbf{v} \in \mathbb{Z}_3^n : d_B(\mathbf{u}, \mathbf{v}) \leq r\} \quad (18)$$

the sphere with center $\mathbf{u}$ and radius $r$ in $\mathbb{Z}_3^n$, and its volume by $|\mathcal{S}(\mathbf{u}, r)|$. The following Proposition gives a lower bound on the value of $|\mathcal{S}(\mathbf{u}, r)|$:

*Proposition 2:* Let $\mathcal{S}(\mathbf{u}, r)$ be a sphere with center $\mathbf{u}$ and radius $r$ in $\mathbb{Z}_3^n$ of volume $|\mathcal{S}(\mathbf{u}, r)|$. It follows that

$$|\mathcal{S}(\mathbf{u}, r)| \geq \sum_{d=0}^{r} \sum_{e_2=0}^{\lfloor d/2 \rfloor} \binom{n}{e_2} \binom{n-e_2}{d-2e_2} \quad (19)$$

where the bound is attained by spheres centered on the vertices of the hypercube.

*Proof:* See Appendix C. ∎

It is now possible to formulate the sphere-packing bound for our channel.

*Theorem 2:* Let $\mathcal{C}_3$ be a ternary code of length $n$ and minimum $d_B$-distance $d_{B,\min}$ over the ternary channel $\mathcal{H}$. It follows that

$$|\mathcal{C}_3| \leq \frac{3^n}{\sum_{d=0}^{t_A} \sum_{e_2=0}^{\lfloor d/2 \rfloor} \binom{n}{e_2} \binom{n-e_2}{d-2e_2}}. \quad (20)$$

*Proof:* See Appendix D. ∎

Note that the tightness of the upper bound in (20) worsens with increasing values of $d_{B,\min}$, since the tightness of the lower bound on the volume of the spheres given by Proposition 2 also decays when $d_{B,\min}$ increases.

## VII. CONSTRUCTIVE LOWER BOUND

In this Section, we give a constructive lower bound on the size of codes over channel $\mathcal{H}$ and show the existence of good codes. Given the code length $n$ and the minimum $d_B$-distance $d_{B,\min}$, the goal is to construct an $[n, M, d_{B,\min}]$ code $\mathcal{C}_3$ for the ternary channel $\mathcal{H}$ with reasonable $M$ and error correcting capability $t_A$ given by (17). The proposed construction method uses binary codes as basic elements. In particular, we define mappings that are applied to binary codes to generate a set of codewords of $\mathbb{Z}_3^n$ that satisfies a given minimum $d_B$-distance $d_{B,\min}$. For clarity purposes, we first summarize the proposed construction method, and then formalize it.

### A. Sketch of the Construction Method

The proposed construction method is a two-step procedure. First, we build a large amount of subspaces of $\mathbb{Z}_3^n$ such that the $d_B$-distance between any two subspaces is at least $d_{B,\min}$[1]. To this aim, we consider an $[n, M_2, d_{B,\min}]$ binary code $\mathcal{C}_2$ for the BSC, such that each codeword $\bar{\mathbf{x}} \in \mathcal{C}_2$ defines a subspace $\mathcal{E}_{\bar{\mathbf{x}}}$ of the ternary space. $\mathcal{E}_{\bar{\mathbf{x}}}$ is the set of ternary words yielding $\bar{\mathbf{x}}$ when projected to binary words (by changing their symbols 2 into symbol 1). Notice that the cardinality of subspaces $\mathcal{E}_{\bar{\mathbf{x}}}$ is $|\mathcal{E}_{\bar{\mathbf{x}}}| = 2^{w_{\bar{\mathbf{x}}}}$.

*Example 1:* Let $\bar{\mathbf{x}} = 1100$. Then,

$$\mathcal{E}_{\bar{\mathbf{x}}} = \{1100, 1200, 2100, 2200\}$$

and $|\mathcal{E}_{\bar{\mathbf{x}}}| = 2^2 = 4$.

The use of codes for the BSC comes from the fact that the binary projection of the transmission chain $(\mathbf{x} \in \mathbb{Z}_3^n) \to \mathcal{H} \to (\mathbf{y} \in \mathbb{Z}_3^n)$ is $(\bar{\mathbf{x}} \in \mathbb{Z}_2^n) \to \bar{\mathcal{H}} \to (\bar{\mathbf{y}} \in \mathbb{Z}_2^n)$, where $\bar{\mathcal{H}}$ is the BSC.

The second step of the code construction is then to select words within every subspace $\mathcal{E}_{\bar{\mathbf{x}}}$ that are distant from each other by at least $d_{B,\min}$. To this end, we consider $\mathcal{E}_{\bar{\mathbf{x}}}$ as a binary space $\mathbb{Z}_2^{w_{\bar{\mathbf{x}}}}$ and use a $[w_{\bar{\mathbf{x}}}, M_2^{w_{\bar{\mathbf{x}}}}, d_{H,\min}]$ code $\mathcal{C}_2^{w_{\bar{\mathbf{x}}}}$ for the binary erasure channel (BEC), with minimum Hamming distance $d_{H,\min} \geq \lceil \frac{d_{B,\min}}{2} \rceil$.

*Example 2:* Let $\bar{\mathbf{x}} = 1100$. Then, $\mathcal{E}_{\bar{\mathbf{x}}}$ is mapped to $\mathbb{Z}_2^2$ by:

$$1100 \to 00, \qquad 1200 \to 01,$$
$$2100 \to 10, \qquad 2200 \to 11.$$

Now, if we choose $\mathcal{C}_2^2 = \{00, 11\}$, then the selected ternary codewords in $\mathcal{E}_{\bar{\mathbf{x}}}$ are 1100 and 2200.

Notice that if $\mathbf{x} \in \mathcal{E}_{\bar{\mathbf{x}}}$ is transmitted, the received vector $\mathbf{y}$ might not belong to $\mathcal{E}_{\bar{\mathbf{x}}}$. If the receiver is able to determine that $\mathbf{x} \in \mathcal{E}_{\bar{\mathbf{x}}}$, i.e., $\hat{\bar{\mathbf{x}}} = \bar{\mathbf{x}}$, we know the position of the zeros of $\mathbf{x}$. Therefore, only errors in the remaining positions (in the form $1 \to 0$ and $2 \to 0$) must be considered. These transitions correspond to erasures in a BEC, hence $\mathcal{C}_2^{w_{\bar{\mathbf{x}}}}$ must be a good code for the BEC able to correct at least $t_A$ errors. Notice that for a BEC this corresponds to $d_{H,\min} \geq t_A + 1 = \lceil \frac{d_{B,\min}}{2} \rceil$.

*Example 3:* Let $\mathbf{x} = 2200$ ($\bar{\mathbf{x}} = 1100$). Assume that $\mathbf{y} = 0200$ was received and that the receiver is able to correctly estimate $\hat{\bar{\mathbf{x}}} = 1100$. If $\mathcal{C}_2^2 = \{00, 11\}$ was chosen, then 0200 is mapped to ?1 where we use symbol ? to denote an erasure. Then, the decoder of $\mathcal{C}_2^2 = \{00, 11\}$ will decode ?1 to 11, which corresponds to the ternary codeword 2200 in $\mathcal{E}_{1100}$.

The set of ternary vectors selected within the subspaces $\mathcal{E}_{\bar{\mathbf{x}}}$ using codes $\mathcal{C}_2^{w_{\bar{\mathbf{x}}}}$ forms the $[n, M, d_{B,\min}]$ ternary code $\mathcal{C}_3$.

---

[1]Here, the distance between two non-empty subsets $\mathcal{S}_1$ and $\mathcal{S}_2$ of a metric space is defined as the minimum distance between any two elements $s_1 \in \mathcal{S}_1$ and $s_2 \in \mathcal{S}_2$.

## B. Mappings and Their Topological Properties

Let $\mathbf{u}$ be a vector of $\mathbb{Z}_2^n$ and denote by $w_\mathbf{u}$ its Hamming weight. We denote by $g_\mathbf{u}(j)$ $(1 \le j \le w_\mathbf{u})$ the $j$th non-zero entry of $\mathbf{u}$. We define the mapping $\varphi_\mathbf{u}$ such that:

$$\varphi_\mathbf{u}: \begin{array}{ccc} \mathbb{Z}_3^{w_\mathbf{u}} & \longrightarrow & \mathbb{Z}_3^n \\ \boldsymbol{a} & \longmapsto & \sum_{j=1}^{w_\mathbf{u}} a_j e_{g_\mathbf{u}(j)} \end{array} \quad (21)$$

where $(e_i)_{1 \le i \le n}$ is the canonical basis of $\mathbb{Z}_3^n$. We also call $\mathcal{E}_\mathbf{u}$ the subspace of $\mathbb{Z}_3^n$ defined by $\mathcal{E}_\mathbf{u} = \varphi_\mathbf{u}(\mathbb{Z}_3^{w_\mathbf{u}})$.

*Example 4:* For $\mathbf{u} = 10011000$, we have $g_\mathbf{u}(1) = 1$, $g_\mathbf{u}(2) = 4$, $g_\mathbf{u}(3) = 5$, and $\varphi_\mathbf{u}(201) = 20001000$. In general, the elements of $\mathcal{E}_\mathbf{u}$ are the vectors of the form $a00bc000$ for $a, b, c \in \mathbb{Z}_3$.

We define another mapping $\psi$ that transforms a word in $(\mathbb{Z}_2 \cup \{?\})^n$ into a ternary word by mapping $0 \to 1$, $1 \to 2$, and $? \to 0$:

$$\psi: \begin{array}{ccc} (\mathbb{Z}_2 \cup \{?\})^n & \longrightarrow & \mathbb{Z}_3^n \\ \boldsymbol{b} & \longmapsto & \sum_{i=1}^n \psi(b_i) e_i \end{array} \quad (22)$$

where

$$\psi(?) = 0, \ \psi(0) = 1, \ \text{and } \psi(1) = 2. \quad (23)$$

*Example 5:* For $\boldsymbol{b} = 11010?1?$, we have $\psi(11010?1?) = 22121020$.

The mappings (21) and (22) have several topological properties regarding $d_\mathrm{B}$:

*Proposition 3:* Let $\mathbf{u}$ and $\mathbf{v}$ be two vectors in $\mathbb{Z}_2^n$, and $\psi$ the mapping defined in (22). It follows that

$$d_\mathrm{B}(\psi(\mathbf{u}), \psi(\mathbf{v})) = 2 d_\mathrm{B}(\mathbf{u}, \mathbf{v}). \quad (24)$$

*Proof:* Since $d_\mathrm{B}(1,2) = 2 d_\mathrm{B}(0,1)$ and $d_\mathrm{B}(a,a) = 0$ for all $a \in \mathbb{Z}_3$, we have

$$d_\mathrm{B}(\psi(\mathbf{u}), \psi(\mathbf{v})) = \sum_{i=1}^n d_\mathrm{B}((\psi(\mathbf{u}))_i, (\psi(\mathbf{v}))_i)$$
$$= \sum_{i=1}^n d_\mathrm{B}(u_i + 1, v_i + 1) \quad (25)$$
$$= \sum_{i=1}^n 2 d_\mathrm{B}(u_i, v_i) = 2 d_\mathrm{B}(\mathbf{u}, \mathbf{v}).$$

∎

*Proposition 4:* Let $\mathbf{u} \in \mathbb{Z}_2^n$. For $\tilde{\mathbf{u}}, \tilde{\mathbf{u}}' \in \mathbb{Z}_3^{w_\mathbf{u}}$, the following equality holds:

$$d_\mathrm{B}(\varphi_\mathbf{u}(\tilde{\mathbf{u}}), \varphi_\mathbf{u}(\tilde{\mathbf{u}}')) = d_\mathrm{B}(\tilde{\mathbf{u}}, \tilde{\mathbf{u}}'). \quad (26)$$

*Proof:*

$$d_\mathrm{B}(\varphi_\mathbf{u}(\tilde{\mathbf{u}}), \varphi_\mathbf{u}(\tilde{\mathbf{u}}')) = d_\mathrm{B}\left( \sum_{j=1}^{w_\mathbf{u}} \tilde{u}_j e_{g_\mathbf{u}(j)}, \sum_{j=1}^{w_\mathbf{u}} \tilde{u}'_j e_{g_\mathbf{u}(j)} \right)$$
$$= \sum_{j=1}^{w_\mathbf{u}} d_\mathrm{B}(\tilde{u}_j, \tilde{u}'_j) = d_\mathrm{B}(\tilde{\mathbf{u}}, \tilde{\mathbf{u}}'). \quad (27)$$

∎

*Proposition 5:* Let $\mathbf{u}, \mathbf{v} \in \mathbb{Z}_2^n$, and let $\tilde{\mathbf{u}} \in \mathbb{Z}_3^{w_\mathbf{u}}$ and $\tilde{\mathbf{v}} \in \mathbb{Z}_3^{w_\mathbf{v}}$, both with no zero entries. The following inequality holds:

$$d_\mathrm{B}(\varphi_\mathbf{u}(\tilde{\mathbf{u}}), \varphi_\mathbf{v}(\tilde{\mathbf{v}})) \ge d_\mathrm{B}(\mathbf{u}, \mathbf{v}). \quad (28)$$

*Proof:* Since $\tilde{\mathbf{u}}$ and $\tilde{\mathbf{v}}$ have no zero entries:

$$d_\mathrm{B}(\varphi_\mathbf{u}(\tilde{\mathbf{u}}), \varphi_\mathbf{v}(\tilde{\mathbf{v}})) \ge d_\mathrm{B}(\varphi_\mathbf{u}(\mathbf{1}_{w_\mathbf{u}}), \varphi_\mathbf{v}(\mathbf{1}_{w_\mathbf{v}})) = d_\mathrm{B}(\mathbf{u}, \mathbf{v}) \quad (29)$$

where $\mathbf{1}_x$ denotes the all-one vector of length $x$. ∎

## C. Construction and Lower Bound

Let $\mathcal{C}_2$ be an $[n, M_2, d_{\mathrm{B,min}}]$ binary code with minimum Hamming distance $d_{\mathrm{H,min}} = d_{\mathrm{B,min}}$ and denote by $A_d$ its weight enumerator (WE), the number of codewords of weight $d$ ($0 \le d \le n$). For all values of $d$ such that $A_d \ne 0$, let $\mathcal{C}_2^d$ be a $[d, M_2^d, d_{\mathrm{H,min}}]$ binary code with $d_{\mathrm{H,min}} \ge \left\lceil \frac{d_{\mathrm{B,min}}}{2} \right\rceil$. We consider the following ternary code:

$$\mathcal{C}_3 = \bigcup_{\bar{\mathbf{x}} \in \mathcal{C}_2} \varphi_{\bar{\mathbf{x}}}(\psi(\mathcal{C}_2^{w_{\bar{\mathbf{x}}}})). \quad (30)$$

*Proposition 6:* The cardinality of code $\mathcal{C}_3$ satisfies:

$$|\mathcal{C}_3| = \sum_{d=0}^n A_d |\mathcal{C}_2^d|. \quad (31)$$

*Proof:* Since for all $\bar{\mathbf{x}} \in \mathcal{C}_2$, $\varphi_{\bar{\mathbf{x}}}$ and $\psi$ are trivially injective, it is enough to prove that the union $\bigcup_{\bar{\mathbf{x}} \in \mathcal{C}_2} \varphi_{\bar{\mathbf{x}}}(\psi(\mathcal{C}_2^{w_{\bar{\mathbf{x}}}}))$ is disjoint.

For $\bar{\mathbf{x}}, \bar{\mathbf{z}} \in \mathcal{C}_2$ such that $\bar{\mathbf{x}} \ne \bar{\mathbf{z}}$, let $\mathbf{x} \in \varphi_{\bar{\mathbf{x}}}(\psi(\mathcal{C}_2^{w_{\bar{\mathbf{x}}}}))$ and $\mathbf{z} \in \varphi_{\bar{\mathbf{z}}}(\psi(\mathcal{C}_2^{w_{\bar{\mathbf{z}}}}))$. By Proposition 5, $d_\mathrm{B}(\mathbf{x}, \mathbf{z}) \ge d_\mathrm{H}(\bar{\mathbf{x}}, \bar{\mathbf{z}}) > 0$, and thus $\mathbf{x} \ne \mathbf{z}$. ∎

*Corollary 1:*

$$|\mathcal{C}_3| = \sum_{d=0}^n A_d M_2^d. \quad (32)$$

*Proposition 7:* Let $\mathbf{x}$ and $\mathbf{z}$ be two distinct codewords of $\mathcal{C}_3$. Then $d_\mathrm{B}(\mathbf{x}, \mathbf{z}) \ge d_{\mathrm{B,min}}$.

*Proof:* See Appendix E. ∎

Therefore, we have constructed an $[n, M, d_{\mathrm{B,min}}]$ ternary code $\mathcal{C}_3$, where $M = \sum_{d=0}^n A_d M_2^d$ (see Proposition 6 above), starting from the binary codes $\mathcal{C}_2$ and $\{\mathcal{C}_2^d\}$ ($0 \le d \le n$).

*Example 6:* We construct a $[5, 21, 3]$ code $\mathcal{C}_3$ for the channel $\mathcal{H}$. First, we consider the binary code $\mathcal{C}_2$ with parameters $[5, 4, 3]$ defined by

$$\mathcal{C}_2 = \{00100, 11000, 00011, 11111\}. \quad (33)$$

Its weight enumerator has three non-zero values: $A_1 = 1$, $A_2 = 2$, and $A_5 = 1$. Therefore, we require three binary codes $\mathcal{C}_2^1$, $\mathcal{C}_2^2$, and $\mathcal{C}_2^5$ of minimum Hamming distance at least $\left\lceil \frac{3}{2} \right\rceil = 2$. We choose $\mathcal{C}_2^1 = \{0\}$, $\mathcal{C}_2^2 = \{00, 11\}$, and $\mathcal{C}_2^5$ the code with generator matrix

$$\begin{pmatrix} 0 & 0 & 0 & 1 & 1 \\ 0 & 0 & 1 & 1 & 0 \\ 0 & 1 & 1 & 0 & 0 \\ 1 & 1 & 0 & 0 & 0 \end{pmatrix}.$$

The code $\mathcal{C}_3$ is obtained by applying (30). The construction of $\mathcal{C}_3$ is represented in Fig. 6. For each codeword $\bar{\mathbf{x}} \in \mathcal{C}_2$ we obtain $|C_2^{w_{\bar{x}}}|$ codewords of $C_3$ through $\varphi_{\bar{x}}(\psi(C_2^{w_{\bar{x}}}))$. For instance, for $\bar{x} = 11000$ and $C_2^2 = \{00, 11\}$ we obtain $|C_2^2| = 2$ codewords of $C_3$ by applying $\varphi_{11000}$ to $\psi(z)$, where $z \in \{00, 11\}$ is one of the codewords of $C_2^2$:

$$\psi(00) = 11, \text{ then } \varphi_{11000}(11) = 11000$$
$$\psi(11) = 22, \text{ then } \varphi_{11000}(22) = 22000.$$

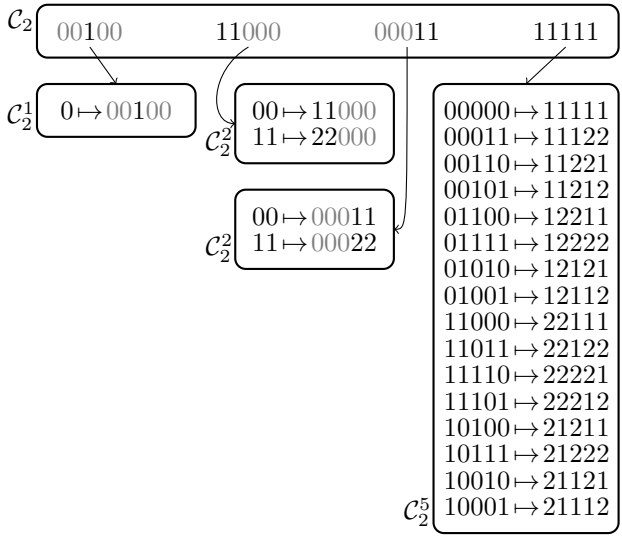

Fig. 6. Example of the construction of a $[5, 21, 3]$ ternary code $C_3$ for channel $\mathcal{H}$ using binary codes as basic elements. The arrows '$\mapsto$' represent the application of the mappings (21) and (22) to the codewords of $C_2$ and to the codewords of the codes $C_2^{w_{\bar{x}}}$, respectively.

TABLE II
SIZE $M$ OF TERNARY CODES OBTAINED USING BINARY CODES FROM [12] AS BASIC ELEMENTS. FOR EACH LENGTH $n$ AND SIZE $M_2$ OF THE CODES $C_2$, WE REPORT THE SIZE $M_{\max}$ (RESP. $M_{\min}$) OF THE LARGEST (RESP. SMALLEST) TERNARY CODE OBTAINED, AND THE AVERAGE SIZE $\bar{M}$. THE NUMBERS BETWEEN BRACKETS ARE THE NUMBER OF BINARY CODES OF SIZE $M_2$.

| $n$ | $M_2$ | | $M_{\max}$ | $M_{\min}$ | $\bar{M}$ |
|---|---|---|---|---|---|
| 7 | 9 | (382) | 120 | 95 | 106 |
|   | 10 | (174) | 124 | 99 | 112 |
|   | 11 | (54) | 129 | 107 | 118 |
|   | 12 | (28) | 133 | 115 | 124 |
|   | 13 | (8) | 137 | 125 | 133 |
|   | 14 | (4) | 141 | 133 | 137 |
|   | 15 | (2) | 145 | 141 | 143 |
|   | 16 | (1) | **149** | 149 | 149 |
| 8 | 18 | (35094) | 307 | 246 | 278 |
|   | 19 | (431) | **311** | 263 | 290 |
|   | 20 | (10) | 309 | 293 | 301 |
| 9 | 36 | (38996) | 835 | 677 | 776 |
|   | 37 | (1464) | 833 | 745 | 792 |
|   | 38 | (116) | **837** | 777 | 806 |
|   | 39 | (6) | 833 | 809 | 817 |
|   | 40 | (2) | 825 | 825 | 825 |
| 10 | 72 | (1124) | **2298** | 2088 | 2204 |
| 11 | 144 | (13088) | **6653** | 6195 | 6586 |

In total, $C_3$ has $A_1|C_2^1| + A_2|C_2^2| + A_5|C_2^5| = 21$ codewords.

*Example 7:* For a comparison with the $[5, 21, 3]$ code from Example 6 above, we tabulate here an optimal $[5, 27, 3]$ code $C_3'$ found by computer search. The code is defined by

$$\begin{aligned}C_3' = \{&01112, 00200, 00121, 01001, 01022, 02110, 10010,\\&02221, 10102, 11111, 11120, 11221, 12020, 12101,\\&12202, 12211, 20020, 20012, 11212, 20111, 21100,\\&21211, 21202, 22001, 22210, 22102, 22222\}.\end{aligned}$$
(34)

The constructive method proposed above gives a lower bound on the cardinality of ternary codes over $\mathcal{H}$. We used this method to construct codes using extended BCH (eBCH) codes for $C_2$ and codes obtained from the tables in [10, 11] for $\{C_2^d\}$. Note that the proposed construction method does not require full knowledge of the binary codes used as basic elements to compute the lower bound; given $n$ and $d_{B,\min}$, only the knowledge of the weight enumerator $A_d$ of $C_2$ is required. On the other hand, for codes $\{C_2^d\}$, only the knowledge of the size $M_2^d$ is required. The results are shown in Table I. For given $n$ and $d_{B,\min}$, we report in the table the code size $M$. The upper bound on the size of codes over $\mathcal{H}$ of length $n$ and minimum $d_B$-distance $d_{B,\min}$ is also given in the table (numbers between brackets).

The constructive lower bound is strongly dependent on the binary codes selected as basic elements. Better bounds than the ones in Table I can be obtained if we use, e.g., good non-linear binary codes instead of eBCH codes. Furthermore, the choice of the binary code $C_2$, even among codes of the same length, minimum distance, and size, can have a great impact on the size of the ternary code generated. We observed that the overall WE of the code was crucial; maximizing the size of $C_2$ may be less important than using a code with a WE adapted to the construction. As an example, for short block lengths $n$ we report in Table II the best size ($M_{\max}$), the worst size ($M_{\min}$), and the average size ($\bar{M}$) of the ternary codes obtained by using (non-linear) binary codes of a given size and minimum distance 3 [12] for $C_2$. The number of codes of a given size $M_2$ is given between brackets in the table, and the largest tabulated value for $M_2$ for a given $n$ is the maximum possible, i.e., the corresponding codes are optimal. For $\{C_2^d\}$ we used the codes from [10, 11], since for minimum distance 2 they are optimal (their size is $M_2^d = 2^{d-1}$). For $n = 11$, there exist 13088 codes of optimal size ($M_2 = 144$). Among them, choosing a code with a more suited WE can bring the size of the generated ternary code from $M_{\min} = 6195$ to $M_{\max} = 6653$. For $n = 9$, the best ternary codes are not obtained from binary codes for $C_2$ of optimal size ($M_2 = 40$), but from a binary code of size $M_2 = 38$. In fact, starting from an optimal code yields worse results than considering codes of smaller sizes, down to $M_2 = 36$.

*Remark 4:* With respect to the list of codes in [12] we also considered the codes obtained by adding (modulo 2) the all-one vector to every codeword of the code, since this changes the WE of the code, which may have a great impact on our construction.

## VIII. VARIABLE-LENGTH TO FIXED-LENGTH ENCODING

The construction method proposed in the previous Section provides codes as sets of codewords. However, finding a simple encoding from the set of messages (thought of as binary words of fixed length) to the set of codewords is a difficult problem. Notice that by construction the resulting ternary codes are non-linear. While it is always possible to enumerate the codewords, for large values of $M$ coding and decoding become far too complex.

TABLE I
CONSTRUCTIVE LOWER BOUND OBTAINED USING eBCH CODES AS BASIC ELEMENTS AND UPPER BOUND (IN BRACKETS) ON THE SIZE $M$ OF TERNARY CODES FOR DIFFERENT VALUES OF THE BLOCK LENGTH $n$ AND MINIMUM $d_\text{B}$-DISTANCE $d_\text{B,min}$.

| $d_\text{B,min} \downarrow$ | $n$ 8 | | 16 | | 32 | | 64 | | 128 | |
|---|---|---|---|---|---|---|---|---|---|---|
| 2 | 3281 | (6561) | 2.15E7 | (4.30E7) | 9.26E14 | (1.85E15) | 1.71E30 | (3.43E30) | 5.89E60 | (1.17E61) |
| 4 | 241 | (729) | 675681 | (2.53E6) | 1.44E13 | (5.61E13) | 1.34E28 | (5.28E28) | 2.30E58 | (9.13E58) |
| 6 | | | 5985 | (281351) | 2.84E10 | (3.30E12) | 6.55E24 | (1.60E27) | 2.81E54 | (1.40E57) |
| 8 | 17 | (41) | 2529 | (45169) | 5.33E8 | (2.84E11) | 5.14E22 | (7.17E25) | 1.09E52 | (3.22E55) |
| 10 | | | | | | | 4.72E19 | (4.23E24) | 1.29E48 | (9.77E53) |
| 12 | | | | | 2.74E6 | (4.43E9) | 3.12E18 | (3.09E23) | 5.17E45 | (3.68E52) |
| 14 | | | | | | | 3.13E15 | (2.68E22) | 2.12E42 | (1.65E51) |
| 16 | | | 33 | (253) | 133057 | (1.37E8) | 1.00E14 | (2.68E21) | 8.22E39 | (8.61E49) |
| 20 | | | | | | | | | 3.42E35 | (3.37E47) |
| 22 | | | | | | | 7.30E10 | (5.40E18) | 8.53E31 | (2.47E46) |
| 24 | | | | | | | 6.92E10 | (8.26E17) | 3.20E29 | (1.98E45) |
| 28 | | | | | | | 1.07E9 | (2.44E16) | 1.00E26 | (1.61E43) |
| 32 | | | | | 65 | (7817) | 2.68E8 | (9.54E14) | 4.25E23 | (1.74E41) |
| 44 | | | | | | | | | 1.84E19 | (8.52E35) |
| 48 | | | | | | | | | 1.15E18 | (2.09E34) |
| 56 | | | | | | | | | 1.12E15 | (2.03E31) |
| 64 | | | | | | | 129 | (5.85E6) | 8.79E12 | (3.39E28) |
| 128 | | | | | | | | | 257 | (2.47E12) |

To circumvent this drawback, we can consider a variable-rate encoding alternative. Consider the $[n, M, d_\text{B,min}]$ ternary code $\mathcal{C}_3$ constructed following the construction method of the previous Section starting from the $[n, M_2, d_\text{B,min}]$ binary code $\mathcal{C}_2$ and the $[d, M_2^d, d_\text{H,min}]$ binary codes $\{\mathcal{C}_2^d\}$ with $d_\text{H,min} \geq \left\lceil \frac{d_\text{B,min}}{2} \right\rceil$. Assume also that efficient encoders and decoders are known for these binary codes over the BSC for $\mathcal{C}_2$ and over the BEC for the codes $\{\mathcal{C}_2^d\}$. Let us define the message $m$ to be transmitted as an infinite sequence of bits. A simple way to progressively encode pieces of $m$ by $\mathcal{C}_3$ is as follows:

1) Let $\mathbf{u}_1$ denote the prefix of length $k$ bits of $m$.
2) Let $\bar{\mathbf{x}}_1$ denote the codeword associated to $\mathbf{u}_1$ by $\mathcal{C}_2$, and $w_{\bar{\mathbf{x}}_1}$ its Hamming weight.
3) We consider the $[w_{\bar{\mathbf{x}}_1}, M_2^{w_{\bar{\mathbf{x}}_1}}, d_\text{H,min}]$ binary code $\mathcal{C}_2^{w_{\bar{\mathbf{x}}_1}}$ with $d_\text{H,min} \geq \lceil d_\text{B,min}/2 \rceil$. Let $k_{w_{\bar{\mathbf{x}}_1}}$ denote the information block length of $\mathcal{C}_2^{w_{\bar{\mathbf{x}}_1}}$, and let $\mathbf{u}_2$ be the next $k_{w_{\bar{\mathbf{x}}_1}}$ bits of $m$ and $\bar{\mathbf{x}}_2$ the codeword associated to $\mathbf{u}_2$ by $\mathcal{C}_2^{w_{\bar{\mathbf{x}}_1}}$.
4) Transmit $\mathbf{x} = \varphi_{\bar{\mathbf{x}}_1}(\psi(\bar{\mathbf{x}}_2))$ over $\mathcal{H}$ and remove the first $k + k_{w_{\bar{\mathbf{x}}_1}}$ bits of $m$.
5) Go back to step 1.

This encoder outputs a sequence of ternary words of length $n$ that are decoded after transmission over $\mathcal{H}$ by a decoder that works with the following pattern:

1) Consider the first received block of $n$ ternary symbols, $\mathbf{y}$.
2) Let $\bar{\mathbf{y}}_1$ denote the word obtained by replacing every occurrence of symbol 2 by the symbol 1 in $\mathbf{y}$.
3) Let $\hat{\mathbf{u}}_1$ denote the output of the decoder of $\mathcal{C}_2$ corresponding to $\bar{\mathbf{y}}_1$ (the estimate of $\mathbf{u}_1$), $\hat{\bar{\mathbf{x}}}_1$ the codeword associated to $\hat{\mathbf{u}}_1$ by $\mathcal{C}_2$, and $w_{\hat{\bar{\mathbf{x}}}_1}$ the weight of $\hat{\bar{\mathbf{x}}}_1$.
4) Let $\bar{\mathbf{y}}_2 = \psi^{-1}\left(\varphi_{\hat{\bar{\mathbf{x}}}_1}^{-1}(\hat{\bar{\mathbf{x}}}_1 * \mathbf{y})\right)$, where $\hat{\bar{\mathbf{x}}}_1 * \mathbf{y}$ denotes the element-wise product of the two vectors.
5) Let $\hat{\mathbf{u}}_2$ denote the output of the decoder $\mathcal{C}_2^{w_{\hat{\bar{\mathbf{x}}}_1}}$ corresponding to $\bar{\mathbf{y}}_2$.
6) Output the concatenation of $\hat{\mathbf{u}}_1$ and $\hat{\mathbf{u}}_2$, and go back to step 1 to decode the next block.

*Proposition 8:* On every packet sent using an $[n, M, d_\text{B,min}]$ code $\mathcal{C}_3$, if less than $t_\text{A} = \left\lfloor \frac{d_\text{B,min}-1}{2} \right\rfloor$ errors occur, the message is correctly decoded.

*Proof:* See Appendix F. ∎

The first drawback of this effective transmission scheme is inherent to the variable-length to fixed-length setting. For finite messages $m$, the length of $m$ will not always match the required information block length of the code. In this situation, some dummy symbols must be appended to the message prior to encoding. While this is not especially a hard problem (a simple solution is to append at the end of the message the symbol 1, and as many symbols 0 as needed to reach the required size, which is easy to code and to decode), it suffers from an efficiency loss that increases as the average number of blocks sent per message decreases.

The second obvious drawback is that if more than $t_\text{A}$ errors occur on the same block, it is very likely that the decoder of $\mathcal{C}_2$ will decode on a codeword of wrong weight, which would result in a shift of the rest of the decoded blocks. The risk of losing such amount of data is affordable only in applications in which any error in the whole message compromises its use, such as the binaries of a software.

## IX. EXTENSION TO $q$-ARY CHANNELS

In this Section, we extend the construction method of Section VII to $q$-ary codes for the $q$-ary generalization of the ternary channel $\mathcal{H}$.

*Definition 5:* For $q \geq 3$, let $\mathcal{H}_q$ be the channel characterized by input alphabet $\mathcal{X} = \{0, \ldots, q-1\}$, output alphabet $\mathcal{Y} = \{0, \ldots, q-1\}$, and the set of conditional probabilities $p(y|x)$ such that for $x \in \mathcal{X}$ and $y \in \mathcal{Y}$:

$$p(y|x) = \begin{cases} 1-p & \text{if } x = y = 0, \\ 1 - \frac{p}{q-1} & \text{if } x = y \neq 0, \\ \frac{p}{q-1} & \text{if } x \neq y \text{ and } xy = 0, \\ 0 & \text{otherwise.} \end{cases} \quad (35)$$

Let $\mathbf{u}$ be a vector of $\mathbb{Z}_2^n$ of Hamming weight $w_\mathbf{u}$. We extend the mappings defined in Section VII to the new $q$-ary setting:

$$\varphi_\mathbf{u}: \quad \mathbb{Z}_q^{w_\mathbf{u}} \longrightarrow \mathbb{Z}_q^n$$
$$\boldsymbol{a} \longmapsto \sum_{j=1}^{w_\mathbf{u}} a_j e_{g_\mathbf{u}(j)} \quad (36)$$

and

$$\psi: \quad (\mathbb{Z}_{q-1} \cup \{?\})^n \longrightarrow \mathbb{Z}_q^n$$
$$\boldsymbol{b} \longmapsto \sum_{i=1}^{n} \psi(b_i) e_i \quad (37)$$

where

$$\psi(?) = 0 \text{ and } \psi(i) = i+1,\ 0 \leq i \leq q-2. \quad (38)$$

We call $\mathcal{E}_\mathbf{u}^q$ the subspace of $\mathbb{Z}_q^n$ defined by $\mathcal{E}_\mathbf{u}^q = \varphi_\mathbf{u}\left(\mathbb{Z}_q^{w_\mathbf{u}}\right)$. These extended mappings maintain the topological properties that they have for the ternary case regarding $d_\mathrm{B}$ (see Section VII-B). Notice that the definition of the distances $d_\mathrm{A}$ and $d_\mathrm{B}$ does not need to be changed to consider the distance between vectors in $\mathbb{Z}_q^n$.

*Proposition 9:* Let $\mathbf{u}$ and $\mathbf{v}$ be two vectors in $\mathbb{Z}_{q-1}^n$. It follows that

$$d_\mathrm{B}(\psi(\mathbf{u}), \psi(\mathbf{v})) = 2d_\mathrm{B}(\mathbf{u}, \mathbf{v}). \quad (39)$$

*Proposition 10:* Let $\mathbf{u} \in \mathbb{Z}_2^n$. For $\tilde{\mathbf{u}}, \tilde{\mathbf{u}}' \in \mathbb{Z}_q^{w_\mathbf{u}}$, the following equality holds:

$$d_\mathrm{B}(\varphi_\mathbf{u}(\tilde{\mathbf{u}}), \varphi_\mathbf{u}(\tilde{\mathbf{u}}')) = d_\mathrm{B}(\tilde{\mathbf{u}}, \tilde{\mathbf{u}}'). \quad (40)$$

*Proposition 11:* Let $\mathbf{u}, \mathbf{v} \in \mathbb{Z}_2^n$, and let $\tilde{\mathbf{u}} \in \mathbb{Z}_q^{w_\mathbf{u}}$ and $\tilde{\mathbf{v}} \in \mathbb{Z}_q^{w_\mathbf{v}}$, both with no zero entries. The following inequality holds:

$$d_\mathrm{B}(\varphi_\mathbf{u}(\tilde{\mathbf{u}}), \varphi_\mathbf{v}(\tilde{\mathbf{v}})) \geq d_\mathrm{B}(\mathbf{u}, \mathbf{v}). \quad (41)$$

Propositions 9 to 11 can be proved in the same way as Propositions 3 to 5 in Section VII-B, respectively.

The goal is now to construct, with $n$ and minimum $d_\mathrm{B}$-distance $d_{\mathrm{B,min}}$ given, an $[n, M, d_{\mathrm{B,min}}]$ code $\mathcal{C}_q$ for the $q$-ary channel $\mathcal{H}_q$ with reasonable $M$, starting from elementary codes for the binary symmetric channel and for the $(q-1)$-ary erasure channel as basic elements.

Let $\mathcal{C}_2$ be an $[n, M_2, d_{\mathrm{B,min}}]$ binary code with minimum Hamming distance $d_{\mathrm{H,min}} = d_{\mathrm{B,min}}$ and denote by $A_d$ its weight enumerator. For all $d$ such that $A_d \neq 0$, let $\mathcal{C}_{q-1}^d$ be a $[d, M_{q-1}^d, d_{\mathrm{H,min}}]$ $(q-1)$-ary code with $d_{\mathrm{H,min}} \geq \lceil d_{\mathrm{B,min}}/2 \rceil$. We consider the $q$-ary code $\mathcal{C}_q$ obtained as follows:

$$\mathcal{C}_q = \bigcup_{\bar{\mathbf{x}} \in \mathcal{C}_2} \varphi_{\bar{\mathbf{x}}}\left(\psi(\mathcal{C}_{q-1}^{w_{\bar{\mathbf{x}}}})\right). \quad (42)$$

*Proposition 12:* The cardinality of code $\mathcal{C}_q$ satisfies:

$$|\mathcal{C}_q| = \sum_{d=0}^{n} A_d |\mathcal{C}_{q-1}^d|. \quad (43)$$

*Corollary 2:*

$$|\mathcal{C}_q| = \sum_{d=0}^{n} A_d M_{q-1}^d. \quad (44)$$

*Proposition 13:* Let $\mathbf{x}$ and $\mathbf{z}$ be two distinct codewords of $\mathcal{C}_q$. Then $d_\mathrm{B}(\mathbf{x}, \mathbf{z}) \geq d_{\mathrm{B,min}}$.

The adaptation of the proofs of these propositions for the ternary case to the $q$-ary case is straightforward, and details are omitted for brevity.

*Remark 5:* The variable-length encoding process in Section VIII for the ternary case can be directly adapted to the $q$-ary case ($q > 3$) when $q$ is of the form $2^\ell + 1$ with $\ell > 1$ by reading symbols of $\mathbb{Z}_{q-1}$ as groups of $\ell$ bits of the binary input. If $q$ is of another form, this direct adaptation is still possible but involves an efficiency loss (the number of bits read is the highest $\ell$ such that $2^\ell + 1 \leq n$, and some symbols of $\mathbb{Z}_{q-1}$ are not used).

## X. TERNARY CODE SEARCH

In this Section, for small values of $n$, we compare the constructive lower bound on the size of ternary codes in Section VII with the results of a computer-based search for good ternary codes. We have conducted both an unrestricted code search for ternary codes and a search based on the code construction from Section VII. As we will show below, in both cases, the code search reduces to the problem of finding (weighted) cliques in an undirected (weighted) graph, which has been solved using state-of-the-art algorithms from graph theory. Both exhaustive algorithms (when the code parameters $n$ and $d_{\mathrm{B,min}}$ are small) and a greedy approximate algorithm (for larger values of the code parameters) have been used.

### A. Unrestricted Ternary Code Search

Let $G = G(V, E, (d_{\mathrm{B,min}}, w_{\min}, w_{\max}))$ denote an undirected graph with vertex set $V$ and edge set $E$, where each vertex $v(\boldsymbol{a})$ represents a ternary vector $\boldsymbol{a} \in \mathbb{Z}_3^n$ with Hamming weight at least $w_{\min}$ and at most $w_{\max}$. Furthermore, $(v(\boldsymbol{a}), v(\boldsymbol{b})) \in E$ if and only if $d_\mathrm{B}(\boldsymbol{a}, \boldsymbol{b}) \geq d_{\mathrm{B,min}}$. Then, an $[n, M]$ ternary block code with codewords of Hamming weight at least $w_{\min}$ and at most $w_{\max}$ and minimum $d_\mathrm{B}$-distance at least $d_{\mathrm{B,min}}$ corresponds to a clique (i.e., a subgraph in which all pairs of vertices are adjacent) of size $M$ in the graph $G$, and a maximum-size (or optimal) $[n, M]$ ternary block code with codewords of Hamming weight at least $w_{\min}$ and at most $w_{\max}$ and minimum $d_\mathrm{B}$-distance at least $d_{\mathrm{B,min}}$ corresponds to a maximum clique in the graph $G$. Thus, the code search problem reduces to finding cliques in an undirected graph.

### B. Restricted Ternary Code Search Based on the Binary Code Construction

Let $G = G(V, E, (d_{\mathrm{B,min}}, w_{\min}, w_{\max}))$ denote an undirected weighted graph with vertex set $V$ and edge set $E$, where each vertex $v(\boldsymbol{a})$ represents a binary vector $\boldsymbol{a} \in \mathbb{Z}_2^n$ with Hamming weight at least $w_{\min}$ and at most $w_{\max}$. Furthermore, the weight of a vertex $v(\mathbf{a})$ is the size of an optimal binary code of length $w_{\boldsymbol{a}}$, where $w_{\boldsymbol{a}}$ denotes the Hamming weight of $\boldsymbol{a}$, and with minimum Hamming distance at least $\left\lceil \frac{d_{\mathrm{B,min}}}{2} \right\rceil$. Also, $(v(\boldsymbol{a}), v(\boldsymbol{b})) \in E$ if and only if $d_\mathrm{B}(\boldsymbol{a}, \boldsymbol{b}) \geq d_{\mathrm{B,min}}$. Then, an $[n, M]$ restricted ternary block code with codewords of Hamming weight at least $w_{\min}$ and at most $w_{\max}$ and minimum $d_\mathrm{B}$-distance at least $d_{\mathrm{B,min}}$ corresponds to a clique of weighted size $M$ in the graph $G$, and a maximum-size (or optimal) restricted $[n, M]$ ternary block code with codewords of Hamming weight at least $w_{\min}$ and at most $w_{\max}$ and minimum $d_{\text{B}}$-distance at least $d_{\text{B,min}}$ corresponds to a maximum weighted clique in the graph $G$. Thus, the code search problem reduces to finding (weighted) cliques in a graph.

Finding the weights of the vertices in the graph above is a difficult problem, since they correspond to the sizes of optimal binary codes. Obviously, we can find these weights by carrying out several code searches in a similar fashion as described above, where the vertex set of the $i$th graph, $i = 1, \ldots, n$, corresponds to a set of binary vectors from $\mathbb{Z}_2^i$. Also, in the $i$th graph, $(v(\boldsymbol{a}), v(\boldsymbol{b})) \in E$ if and only if $d_{\text{H}}(\boldsymbol{a}, \boldsymbol{b}) \geq \left\lceil \frac{d_{\text{B,min}}}{2} \right\rceil$.

### C. Exhaustive Search for Maximum Cliques

The problem of finding a maximum clique (or the equivalent problem of finding a maximum stable set) in an arbitrary undirected graph is one of the most important NP-hard problems in discrete mathematics and theoretical computer science. There are several efficient algorithms for searching for maximum cliques in arbitrary graphs. See, for instance, [13–15] and references therein. One standard approach for finding a maximum clique is based on the branch-and-bound method [16]. Most branch-and-bound algorithms use heuristic coloring algorithms to find an upper bound on the size of a maximum clique in the bound step. Sophisticated coloring algorithms can reduce the search space significantly. For very dense graphs (or very sparse graphs for the equivalent problem of finding a maximum stable set), the branch-and-bound algorithm in [14] is one of the fastest known algorithms, and we have used this algorithm in combination with the algorithm from [13], which is faster when the graph is not that dense, when searching for ternary block codes.

Finally, we remark that most algorithms for finding a maximum clique can be straightforwardly extended to finding a maximum weighted clique.

### D. Greedy Search for Maximum Cliques

There are numerous heuristic or approximate algorithms in the literature for searching for maximum (weighted) cliques in an arbitrary (weighted) undirected graph. See, for instance, [17] and references therein. In this work, we have used a very simple greedy algorithm, which is outlined below in Algorithm 1.

The greedy algorithm in Algorithm 1 is a random algorithm, since there could be several vertices $v$ in line 8 that have maximum degree. Thus, it is beneficial to run the algorithm several times.

In the setting of a weighted graph, priority is given to the vertex of maximum degree (as for the unweighted case), and if there are several vertices with degree equal to the maximum degree, priority is given (in a random fashion) to the vertices with the highest weight.

### E. Results From a Code Search

When $n$ is not very small, the size of the unrestricted graph described in Section X-A becomes very large. Thus, to reduce

**Algorithm 1** Greedy Maximum Clique Search

1: /∗ Find an approximate maximum clique in an arbitrary undirected graph $G = G(V, E)$ ∗/
2: Construct the complement $\bar{G} = \bar{G}(V, \bar{E})$ of $G$, where an edge $e \in \bar{E}$ if and only if $e \notin E$.
3: Initialize the set $\tilde{V}$ with the empty set.
4: **while** $V \neq \emptyset$ **do**
5:   **if** $\exists v \in V$ of degree at most 1 **then**
6:     select a random vertex $v \in V$ of degree at most 1 and add it to $\tilde{V}$, i.e., let $\tilde{V} \leftarrow \tilde{V} \cup \{v\}$
7:   **else**
8:     select a random vertex $v \in V$ of maximum degree ($> 1$) and add it to $\tilde{V}$, i.e., let $\tilde{V} \leftarrow \tilde{V} \cup \{v\}$.
9:   **end if**
10:   Remove $v$ and all its adjacent edges from $\bar{G}$.
11: **end while**
12: The vertex set $\tilde{V}$ is an independence set in the original undirected graph $\bar{G}$, and it follows that $\tilde{V}$ is a clique in the original undirected graph $G$.

the size of the graph, we have, in some cases, restricted the code search to constant-weight codes, i.e., to codes in which all codewords have the same Hamming weight ($w_{\min} = w_{\max} = w$, for some $w$), or to nearly constant-weight codes, where $w_{\min}$ is larger than 0 and/or $w_{\max}$ is less than $n$, and $w_{\max} - w_{\min}$ is small compared to $n$. Also, when the size of the graph becomes too large, the greedy algorithm from Algorithm 1 is used instead of a much more complex exhaustive algorithm.

In Table III, the size $M$ of both restricted and unrestricted ternary codes for different values of the block length $n$ and the minimum distance $d_{\text{B,min}}$ are presented. The numbers in the parentheses are from an unrestricted search for ternary codes, and should be compared to the numbers in front that are from a restricted search, as described in Section X-B. The numbers in bold are exact values (from an exhaustive search) for non-constant-weight codes, i.e., with $w_{\min} = 0$ and $w_{\max} = n$. As can be seen from the table, in some cases, the binary code construction gives optimal ternary codes.

Notice that the output of this ternary code search is not purely numerical and actually yields codes as sets of codewords. While this may not always be a valuable information (especially for codes with no known form of regularity), in our case it provides binary codes $C_2$ that yield good or even optimal ternary codes. It also provides the families $C_2^d$, but only the size of these matter, so the problem of finding the best $C_2^d$ is the same as the one of finding the largest binary codes of length $d$, for a given minimum distance.

## XI. CONCLUSION

In this paper, coding for a non-symmetric ternary channel where some transitions are not allowed was addressed. We derived the ML decoding rule for this channel and showed that it is complex to compute, since it depends on the error transition probability $p$. We then proposed an alternative decoding rule, called $d_{\text{A}}$-decoding, based on a more suitable distance measure which does only depend on code properties.

TABLE III
CONSTRUCTIVE LOWER BOUND ON THE SIZE $M$ OF TERNARY CODES FOR DIFFERENT VALUES OF THE BLOCK LENGTH $n$ AND MINIMUM $d_B$-DISTANCE $d_{B,\min}$ WHEN USING A CONSTRUCTION BASED ON BINARY CODES COMBINED WITH A BINARY CODE SEARCH. THE NUMBERS IN THE PARENTHESES ARE FROM AN UNRESTRICTED SEARCH FOR TERNARY CODES, AND THE NUMBERS IN BOLD ARE EXACT VALUES (EXHAUSTIVE SEARCH).

| $d_{B,\min} \downarrow$ | 5 | | 6 | | 7 | | 8 | | 9 | | 10 | 11 |
|---|---|---|---|---|---|---|---|---|---|---|---|---|
| 2 | **122** | (122) | **365** | (365) | **1094** | (1094) | **3281** | (3281) | 9842 | (9842) | 29525 | 88574 |
| 3 | **21** | (27) | **54** | (61) | **149** | (168) | **337** | (383) | 937 | (990) | 2306 | 6581 |
| 4 | **17** | (17) | **38** | (40) | **92** | (94) | **241** | (272) | 545 | (607) | 1482 | 3476 |
| 5 | **5** | (7) | **9** | (14) | **17** | (26) | **25** | (53) | 50 | (117) | **106** | 277 |
| 6 | | | **9** | (12) | **17** | (18) | **21** | (35) | 46 | (77) | **82** | 188 |
| 7 | | | | | **9** | (9) | **17** | (17) | **21** | (25) | **41** | 77 |
| 8 | | | | | | | **17** | (17) | **21** | (21) | **41** | 73 |
| 9 | | | | | | | | | **7** | (11) | **13** | 25 |
| 10 | | | | | | | | | | | **13** | 25 |
| 11 | | | | | | | | | | | | **13** |

We showed that under certain conditions $d_A$-decoding and ML decoding rules are equivalent. Further, we analyzed error correcting capabilities of ternary codes over this particular channel under $d_A$-decoding. We derived an upper bound and a constructive lower bound on the code size, showing the existence of good codes. Following the proposed constructive method, we found good codes for several values of $n$ and $d_{B,\min}$. The proposed construction method was also extended to $q$-ary generalizations of the non-symmetric ternary channel. Finally, the constructive lower bound was compared with results from a clique-based search for optimal ternary codes for small code lengths. It is shown that in some cases the proposed construction method gives optimal codes.

## APPENDIX A
## PROOF OF THEOREM 1

We prove the direct implication and its converse.

- *Direct implication:*
  We assume that (13) does not hold. For $n$ odd, we write $n = 2m+1$. Consider the code $\mathcal{C} = \{\mathbf{0}_n, \mathbf{1}_n\}$ consisting of two codewords, the all-zero codeword and the all-one codeword, and the received vector $\mathbf{y} = \mathbf{0}_{m+1}\mathbf{1}_m$ consisting of $m+1$ zeros and $m$ ones. Clearly, $d_A$-decoding decodes $\mathbf{y}$ to the all-zero codeword $\mathbf{0}_n$. On the other hand,

$$p(\mathbf{y}|\mathbf{0}_n) = (1-p)^{m+1}(p/2)^m \, , \\ p(\mathbf{y}|\mathbf{1}_n) = (p/2)^{m+1}(1-p/2)^m. \quad (45)$$

Using the hypothesis, we obtain $p(\mathbf{y}|\mathbf{0}_n) \leq p(\mathbf{y}|\mathbf{1}_n)$. Therefore, ML decoding will not necessarily decode to $\mathbf{0}_n$.

If $n$ is even, we use the same argument considering the same vectors with an extra zero appended at the end.

- *Converse:*
  We consider a word $\mathbf{y} \in \mathbb{Z}_3^n$ of weight $w_{\mathbf{y}}$. For a given $d$ and a codeword $\mathbf{x}$ such that $d_A(\mathbf{x}, \mathbf{y}) = d$, the conditional probability that $\mathbf{y}$ was received knowing that $\mathbf{x}$ was transmitted is

$$p(\mathbf{y}|\mathbf{x}) = \prod_{i=1}^{n} p(y_i|x_i) \\ = \left(\frac{p}{2}\right)^d (1-p)^{\alpha_0} \left(1-\frac{p}{2}\right)^{n-d-\alpha_0} \quad (46)$$

where $\alpha_0 = |\{i : x_i = y_i = 0\}|$.

Because $1-p \leq 1-\frac{p}{2}$, if $\mathbf{x}$ varies with $n$, $w_{\mathbf{y}}$, and $d$ fixed, the probability (46) decreases for increasing values of $\alpha_0$.

Notice that $\alpha_0$ satisfies $\alpha_0 \leq \min\{n-d, n-w_\mathbf{y}\}$, where $n-d$ is the number of symbols that are equal in $\mathbf{x}$ and $\mathbf{y}$, and $n-w_\mathbf{y}$ is the number of symbols 0 in $\mathbf{y}$. Therefore, $p(\mathbf{y}|\mathbf{x})$, for given $n$, $w_\mathbf{y}$, and $d$, can be lower-bounded by

$$p(\mathbf{y}|\mathbf{x}) \geq B^-(n, w_\mathbf{y}, d)$$
$$= \begin{cases} \left(\frac{p}{2}\right)^d \left(1 - \frac{p}{2}\right)^{w_\mathbf{y}-d} (1-p)^{n-w_\mathbf{y}} & \text{if } d < w_\mathbf{y}, \\ \left(\frac{p}{2}\right)^d (1-p)^{n-d} & \text{otherwise.} \end{cases} \quad (47)$$

Similarly, since $\alpha_0 \geq \max\{0, n-w_\mathbf{y}-d\}$, for given $n$, $w_\mathbf{y}$, and $d$, $p(\mathbf{y}|\mathbf{x})$ can be upper-bounded by

$$p(\mathbf{y}|\mathbf{x}) \leq B^+(n, w_\mathbf{y}, d)$$
$$= \begin{cases} \left(\frac{p}{2}\right)^d \left(1 - \frac{p}{2}\right)^{w_\mathbf{y}} (1-p)^{n-w_\mathbf{y}-d} & \text{if } d < n - w_\mathbf{y}, \\ \left(\frac{p}{2}\right)^d \left(1 - \frac{p}{2}\right)^{n-d} & \text{otherwise.} \end{cases} \quad (48)$$

We now prove that if (13) holds, then $B^+(n, w_\mathbf{y}, d+1) < B^-(n, w_\mathbf{y}, d)$ for all $n$, $w_\mathbf{y}$, and $d$. Thus, we prove that the closest codeword to a received vector $\mathbf{y}$ is always the most likely one. For the sake of simplicity, we will denote these two bounds by $B^+_{d+1} = B^+(n, w_\mathbf{y}, d+1)$ and $B^-_d = B^-(n, w_\mathbf{y}, d)$, respectively. There are 4 subcases depending on the value of $d$ compared to $w_\mathbf{y}$ and to $n - w_\mathbf{y}$:

- For $\begin{cases} d & < & w_\mathbf{y} \\ d+1 & < & n - w_\mathbf{y} \end{cases}$ :

$$\begin{cases} B^-_d = \left(\frac{p}{2}\right)^d \left(1-\frac{p}{2}\right)^{w_\mathbf{y}-d}(1-p)^{n-w_\mathbf{y}}, \\ B^+_{d+1} = \left(\frac{p}{2}\right)^{d+1} \left(1-\frac{p}{2}\right)^{w_\mathbf{y}} (1-p)^{n-d-w_\mathbf{y}-1}. \end{cases}$$

Thus, $B^+_{d+1} < B^-_d$ if and only if

$$\frac{p/2}{1-p} < \left(\frac{1-p}{1-p/2}\right)^d. \quad (49)$$

To prove (49), it is sufficient to show that $d \leq \lfloor \frac{n-1}{2} \rfloor$, because by hypothesis

$$\frac{p/2}{1-p} < \left(\frac{1-p}{1-p/2}\right)^{\lfloor \frac{n-1}{2} \rfloor} \quad (50)$$

and because

$$\frac{1-p}{1-p/2} \leq 1. \quad (51)$$

In this subcase of the proof

$$\begin{cases} d & < & n - w_\mathbf{y} - 1, \\ d & < & w_\mathbf{y}. \end{cases} \quad (52)$$

Therefore, $2d < n - 1$, and $d \leq \lfloor \frac{n-1}{2} \rfloor$.

- For $\begin{cases} d & < & w_\mathbf{y} \\ d+1 & \geq & n - w_\mathbf{y} \end{cases}$ :

$$\begin{cases} B^-_d = \left(\frac{p}{2}\right)^d \left(1-\frac{p}{2}\right)^{w_\mathbf{y}-d}(1-p)^{n-w_\mathbf{y}}, \\ B^+_{d+1} = \left(\frac{p}{2}\right)^{d+1} \left(1-\frac{p}{2}\right)^{n-d-1}. \end{cases}$$

Thus, $B^+_{d+1} < B^-_d$ if and only if

$$\frac{p/2}{1-p} < \left(\frac{1-p}{1-p/2}\right)^{n-w_\mathbf{y}-1}. \quad (53)$$

The proof of (53) is similar to the proof of (49): it is sufficient to show that $n - w_\mathbf{y} - 1 \leq \lfloor \frac{n-1}{2} \rfloor$. Since in this subcase

$$\begin{cases} n - w_\mathbf{y} & \leq & d+1, \\ n - w_\mathbf{y} & \leq & n-d-1, \end{cases} \quad (54)$$

we have $2(n - w_\mathbf{y}) \leq n$, and $n - w_\mathbf{y} - 1 \leq \frac{n}{2} - 1 \leq \lfloor \frac{n-1}{2} \rfloor$.

- For $\begin{cases} d & \geq & w_\mathbf{y} \\ d+1 & < & n - w_\mathbf{y} \end{cases}$ :

$$\begin{cases} B^-_d = \left(\frac{p}{2}\right)^d (1-p)^{n-d}, \\ B^+_{d+1} = \left(\frac{p}{2}\right)^{d+1} \left(1-\frac{p}{2}\right)^{w_\mathbf{y}} (1-p)^{n-d-w_\mathbf{y}-1}. \end{cases}$$

Thus, $B^+_{d+1} < B^-_d$ if and only if

$$\frac{p/2}{1-p} < \left(\frac{1-p}{1-p/2}\right)^{w_\mathbf{y}}. \quad (55)$$

Again, to prove (55), we show that $w_\mathbf{y} \leq \lfloor \frac{n-1}{2} \rfloor$. Now,

$$\begin{cases} w_\mathbf{y} & \leq & n-d-2, \\ w_\mathbf{y} & \leq & d, \end{cases} \quad (56)$$

so we have $2w_\mathbf{y} \leq n - 2$, and $w_\mathbf{y} \leq \lfloor \frac{n-1}{2} \rfloor$.

- For $\begin{cases} d & \geq & w_\mathbf{y} \\ d+1 & \geq & n - w_\mathbf{y} \end{cases}$ :

$$\begin{cases} B^-_d = \left(\frac{p}{2}\right)^d (1-p)^{n-d}, \\ B^+_{d+1} = \left(\frac{p}{2}\right)^{d+1} \left(1-\frac{p}{2}\right)^{n-d-1}. \end{cases}$$

Thus, $B^+_{d+1} < B^-_d$ if and only if

$$\frac{p/2}{1-p} < \left(\frac{1-p}{1-p/2}\right)^{n-d-1}. \quad (57)$$

Again, to prove (57), we show that $n - d - 1 \leq \lfloor \frac{n-1}{2} \rfloor$. As in this subcase,

$$\begin{cases} n - d & \leq & w_\mathbf{y} + 1, \\ n - d & \leq & n - w_\mathbf{y}, \end{cases} \quad (58)$$

we have $2(n-d) \leq n+1$, and $n - d - 1 \leq \lfloor \frac{n-1}{2} \rfloor$.

## APPENDIX B
## PROOF OF PROPOSITION 1

Let $\mathbf{x}$ be a codeword of $\mathcal{C}_3$ transmitted over the ternary channel $\mathcal{H}$, and let $\mathbf{y}$ be the received vector at the output of the channel. If a decoder implementing the $d_A$-decoding rule erroneously decodes $\mathbf{y}$ to $\hat{\mathbf{x}} \neq \mathbf{x}$, then

$$d_A(\mathbf{x}, \mathbf{y}) \geq d_A(\hat{\mathbf{x}}, \mathbf{y}). \quad (59)$$

Using (59) and Definition 3,

$$2d_A(\mathbf{x}, \mathbf{y}) \geq d_A(\mathbf{x}, \mathbf{y}) + d_A(\mathbf{y}, \hat{\mathbf{x}}) \geq d_B(\mathbf{x}, \hat{\mathbf{x}})$$
$$\geq d_{B,\min} > 2 \left\lfloor \frac{d_{B,\min} - 1}{2} \right\rfloor. \quad (60)$$

Therefore, we successfully $d_A$-decode $\mathbf{y}$ if

$$d_A(\mathbf{x}, \mathbf{y}) \leq \left\lfloor \frac{d_{B,\min} - 1}{2} \right\rfloor, \quad (61)$$

where $d_A(\mathbf{x}, \mathbf{y})$ is the number of errors that occurred during the transmission of $\mathbf{x}$.

Conversely, by Definitions 3 and 4, there exist two codewords $\mathbf{x}$ and $\hat{\mathbf{x}}$ and a vector $\mathbf{y} \in \mathbb{Z}_3^n$ such that

$$d_B(\mathbf{x}, \hat{\mathbf{x}}) = d_{B,\min} \\ d_B(\mathbf{x}, \hat{\mathbf{x}}) = d_A(\mathbf{x}, \mathbf{y}) + d_A(\mathbf{y}, \hat{\mathbf{x}}). \quad (62)$$

Therefore, if $d_A(\mathbf{x}, \mathbf{y}) > t_A$, then

$$d_A(\mathbf{x}, \mathbf{y}) \geq \frac{d_{B,\min}}{2} = \frac{d_A(\mathbf{x}, \mathbf{y}) + d_A(\mathbf{y}, \hat{\mathbf{x}})}{2}. \quad (63)$$

Thus, $d_A(\mathbf{y}, \hat{\mathbf{x}}) \leq d_A(\mathbf{x}, \mathbf{y})$, and the $d_A$-decoder may fail to decode $\mathbf{y}$ to $\mathbf{x}$.

## APPENDIX C
## PROOF OF PROPOSITION 2

We first prove that the smallest spheres are the ones centered on words of maximum weight (the vertices of the hypercube).

Let $n$ and $r$ be two integers. For $w \leq n$, let $\mathbf{u}_w$ be a vector of $\mathbb{Z}_3^n$ of weight $w$. The volume of $\mathcal{S}(\mathbf{u}_w, r)$ is independent from the choice of $\mathbf{u}_w$. We denote it by $\mathcal{V}(n, w, r)$. For $n > 0$, we denote by $\mathbf{u}'_w$ the vector of $\mathbb{Z}_3^{n-1}$ obtained by removing the last symbol of $\mathbf{u}_w$:

$$\begin{aligned} \mathcal{V}(n, w, r) &= |\{\mathbf{v} \in \mathbb{Z}_3^n : d_B(\mathbf{u}_w, \mathbf{v}) \leq r\}| \\ &= |\{\mathbf{w}0 : \mathbf{w} \in \mathbb{Z}_3^{n-1} \wedge d_B(\mathbf{u}'_w, \mathbf{w}) \leq r\}| \\ &+ |\{\mathbf{w}1 : \mathbf{w} \in \mathbb{Z}_3^{n-1} \wedge d_B(\mathbf{u}'_w, \mathbf{w}) \leq r-1\}| \\ &+ |\{\mathbf{w}2 : \mathbf{w} \in \mathbb{Z}_3^{n-1} \wedge d_B(\mathbf{u}'_w, \mathbf{w}) \leq r-1\}| \\ &= \mathcal{V}(n-1, w, r) + 2\mathcal{V}(n-1, w, r-1), \end{aligned} \quad (64)$$

where for $\mathbf{w} \in \mathbb{Z}_3^{n-1}$, $\mathbf{w}0$ denotes the vector of $\mathbb{Z}_3^n$ obtained by appending a 0 at the end of $\mathbf{w}$.

Similarly, we show that for $w \leq n-1$,

$$\begin{aligned} \mathcal{V}(n, w+1, r) &= \mathcal{V}(n-1, w, r) + \mathcal{V}(n-1, w, r-1) \\ &+ \mathcal{V}(n-1, w, r-2). \end{aligned} \quad (65)$$

Therefore, if $n > 0$ and $w \leq n-1$,

$$\begin{aligned} \mathcal{V}(n, w, r) &- \mathcal{V}(n, w+1, r) \\ &= \mathcal{V}(n-1, w, r-1) - \mathcal{V}(n-1, w, r-2) \geq 0. \end{aligned} \quad (66)$$

From (66), it follows that the spheres of minimal volume are the ones centered on words of maximum weight.

Now, we give an expression for $\mathcal{V}(n, n, r)$. We consider the all-one vector $\mathbf{1}_n$. Let $\mathbf{v} \in \mathbb{Z}_3^n$: $\mathbf{v}$ is in $\mathcal{S}(\mathbf{1}_n, r)$ if and only if $d_B(\mathbf{1}_n, \mathbf{v}) \leq r$. We denote by $d$ this distance, and by $e_2$ the number of positions $i$ where $v_i = 2$. The number of positions $j$ where $v_j = 0$ is $d - 2e_2$. The number of vectors $\mathbf{v}$ that match a given $d$ and $e_2$ is therefore:

$$\binom{n}{e_2}\binom{n-e_2}{d-2e_2}.$$

We conclude by summing over all possible $d$ and $e_2$:

$$|\mathcal{S}(\mathbf{1}^n, r)| = \sum_{d=0}^{r} \sum_{e_2=0}^{\lfloor d/2 \rfloor} \binom{n}{e_2}\binom{n-e_2}{d-2e_2}. \quad (67)$$

## APPENDIX D
## PROOF OF THEOREM 2

Let $\mathbf{x}$ and $\tilde{\mathbf{x}}$ be two codewords of $\mathcal{C}_3$. Since $d_B(\mathbf{x}, \tilde{\mathbf{x}}) \geq 2t_A + 1$, the spheres $\mathcal{S}(\mathbf{x}, t_A)$ and $\mathcal{S}(\tilde{\mathbf{x}}, t_A)$ are non-intersecting. This implies that

$$\begin{aligned} \left| \bigcup_{\mathbf{x} \in \mathcal{C}_3} \mathcal{S}(\mathbf{x}, t_A) \right| &= \sum_{\mathbf{x} \in \mathcal{C}_3} |\mathcal{S}(\mathbf{x}, t_A)| \\ &\geq |\mathcal{C}_3| \sum_{d=0}^{t_A} \sum_{e_2=0}^{\lfloor d/2 \rfloor} \binom{n}{e_2}\binom{n-e_2}{d-2e_2}. \end{aligned} \quad (68)$$

Furthermore,

$$\left| \bigcup_{\mathbf{x} \in \mathcal{C}_3} \mathcal{S}(\mathbf{x}, t_A) \right| \leq |\mathbb{Z}_3^n| = 3^n. \quad (69)$$

Therefore, we conclude that

$$|\mathcal{C}_3| \leq \frac{3^n}{\sum_{d=0}^{t_A} \sum_{e_2=0}^{\lfloor d/2 \rfloor} \binom{n}{e_2}\binom{n-e_2}{d-2e_2}}.$$

## APPENDIX E
## PROOF OF PROPOSITION 7

Let $\mathbf{x}$ and $\mathbf{z}$ be two distinct codewords of $\mathcal{C}_3$. We denote by $\bar{\mathbf{x}}$ the codeword of $\mathcal{C}_2$ and by $\bar{\mathbf{x}}'$ the codeword of $\mathcal{C}_2^{w_{\bar{\mathbf{x}}}}$ such that $\mathbf{x} = \varphi_{\bar{\mathbf{x}}}(\psi(\bar{\mathbf{x}}'))$ (the unicity is proved in the proof of Proposition 6). Likewise, we define $\bar{\mathbf{z}}$ and $\bar{\mathbf{z}}'$ with respect to $\mathbf{z}$. We consider two cases:

- *Case $\bar{\mathbf{x}} = \bar{\mathbf{z}}$*: In this case, $\bar{\mathbf{x}}'$ and $\bar{\mathbf{z}}'$ are two different codewords of $\mathcal{C}_2^{w_{\bar{\mathbf{x}}}}$ (otherwise $\mathbf{x} = \mathbf{z}$). Thus, by choice of the code $\mathcal{C}_2^{w_{\bar{\mathbf{x}}}}$ it follows that $d_H(\bar{\mathbf{x}}', \bar{\mathbf{z}}') \geq \lceil d_{B,\min}/2 \rceil$. By Propositions 3 and 4,

$$\begin{aligned} d_B(\mathbf{x}, \mathbf{z}) &= d_B(\varphi_{\bar{\mathbf{x}}}(\psi(\bar{\mathbf{x}}')), \varphi_{\bar{\mathbf{z}}}(\psi(\bar{\mathbf{z}}'))) = d_B(\psi(\bar{\mathbf{x}}'), \psi(\bar{\mathbf{z}}')) \\ &= 2d_H(\bar{\mathbf{x}}', \bar{\mathbf{z}}') \geq 2\lceil d_{B,\min}/2 \rceil \geq d_{B,\min}. \end{aligned} \quad (70)$$

- *Case $\bar{\mathbf{x}} \neq \bar{\mathbf{z}}$*: By choice of $\mathcal{C}_2$ it follows that $d_B(\bar{\mathbf{x}}, \bar{\mathbf{z}}) \geq d_{B,\min}$. Now, by Proposition 5,

$$\begin{aligned} d_B(\mathbf{x}, \mathbf{z}) &= d_B(\varphi_{\bar{\mathbf{x}}}(\psi(\bar{\mathbf{x}}')), \varphi_{\bar{\mathbf{z}}}(\psi(\bar{\mathbf{z}}'))) \geq d_B(\bar{\mathbf{x}}, \bar{\mathbf{z}}) \\ &= d_{B,\min}. \end{aligned} \quad (71)$$

In both cases, $d_B(\mathbf{x}, \mathbf{z}) \geq d_{B,\min}$, which concludes the proof.

## APPENDIX F
## PROOF OF PROPOSITION 8

We prove that, if $d(\mathbf{x}, \mathbf{y}) \leq t_A$, then $\mathbf{y}$ is properly decoded, i.e., $\hat{\mathbf{u}}_1 = \mathbf{u}_1$ and $\hat{\mathbf{u}}_2 = \mathbf{u}_2$.

- $\hat{\mathbf{u}}_1 = \mathbf{u}_1$:

$$\begin{aligned} d(\bar{\mathbf{x}}_1, \bar{\mathbf{y}}_1) &= \sum_{i=1}^{n} d(\bar{x}_{1i}, \bar{y}_{1i}) \\ &= |\{i : \bar{x}_{1i} \neq \bar{y}_{1i}\}| \\ &= |\{i : x_i \neq y_i\}| \\ &= d_B(\mathbf{x}, \mathbf{y}) \leq t_A \end{aligned} \quad (72)$$

Therefore, since $\bar{\mathbf{x}}_1$ is a codeword of $\mathcal{C}_2$ which has an error correction capability $t_A$, $\mathbf{y}$ is successfully decoded to $\hat{\mathbf{u}}_1 = \mathbf{u}_1$.

- $\hat{\mathbf{u}}_2 = \mathbf{u}_2$:
  Since $\hat{\mathbf{u}}_1 = \mathbf{u}_1$, we have $\hat{\bar{\mathbf{x}}}_1 = \bar{\mathbf{x}}_1$ and $w_{\hat{\bar{\mathbf{x}}}_1} = w_{\bar{\mathbf{x}}_1}$.
  We prove that $\bar{\mathbf{y}}_2$ can be decoded to $\mathbf{u}_2$ using the decoder of code $\mathcal{C}_2^{w_{\bar{\mathbf{x}}_1}}$. Consider a position $j$, $1 \leq j \leq w_{\bar{\mathbf{x}}_1}$:
  - Either $\bar{y}_{2j} = \bar{x}_{2j}$,
  - Or $\bar{y}_{2j} \neq \bar{x}_{2j}$ and $\bar{y}_{2j} = ?$, which involves that $\hat{\bar{\mathbf{x}}}_1 * \mathbf{y}$ has a 0 at coordinate $i = g_{\hat{\bar{\mathbf{x}}}_1}(j)$ (the $j$th non-zero entry of $\hat{\bar{\mathbf{x}}}_1$). Since $\hat{\bar{x}}_{1i} = 1$, $y_i = 0$. Also, since $\hat{\mathbf{u}}_1 = \mathbf{u}_1$, then $\hat{\bar{\mathbf{x}}}_1 = \bar{\mathbf{x}}_1$, which implies that $g_{\hat{\bar{\mathbf{x}}}_1}(j) = g_{\bar{\mathbf{x}}_1}(j) = i$. Thus, $x_i \neq 0$ and $x_i \neq y_i$, with $y_i = 0$ since $\bar{y}_{2j} = ?$. By assumption there are at most $t_A$ coordinates such that $x_i \neq y_i$. Therefore, we conclude that $|\{j : \bar{y}_{2j} = ?\}| \leq t_A$,
  - Notice that the case $\bar{y}_{2j} \neq \bar{x}_{2j}$ and $\bar{y}_{2j} \neq ?$ is not possible. Following the reasoning of the previous case, it would involve that $\mathbf{x}$ and $\mathbf{y}$ are different but both non-zero at a given coordinate, which is not a possible transition in our channel $\mathcal{H}$.

Therefore, no error and at most $t_A = \left\lfloor \frac{d_{B,\min}-1}{2} \right\rfloor$ erasures differentiate $\bar{\mathbf{y}}_2$ from $\bar{\mathbf{x}}_2$. Since the minimum distance of $\mathcal{C}_2^{w_{\bar{\mathbf{x}}_1}}$ is such that $\left\lceil \frac{d_{B,\min}}{2} \right\rceil > t_A$, its decoder can correct these at most $t_A$ erasures, and successfully estimate $\mathbf{u}_2$.